\documentclass[12pt]{report}
\usepackage[margin=2.5cm]{geometry}
\usepackage{pslatex} 
\newcommand{\bm}[1]{\boldsymbol{#1}} 
\usepackage{graphicx}
\usepackage{latexsym}
\usepackage{amsmath}
\usepackage{amsfonts}
\usepackage{amssymb}
\usepackage{amsthm}
\usepackage{setspace}
\usepackage{tikz} 
\usepackage{ctable}
\usepackage{siunitx} 
\usepackage{pdflscape}
\usepackage{url,apacite}\usepackage{apacite} 

\renewcommand{\APACrefnote}[1]{
  \ifx\@empty#1\@empty
  \else
    {#1}%
  \fi
}
\usepackage{ragged2e}
 \RaggedRight
\usepackage{appendix}
\usepackage{fancyhdr}
\newcommand{\blue}[1]{#1}

{
    \fancyhead[L]{CONFIDENCE SETS AND FUNGIBLE ESTIMATES}
    \fancyhead[R]{\thepage}
    \fancyfoot[C]{}

    \footskip=36pt
    \fancypagestyle{title}{
        \fancyhead[L]{Running head: CONFIDENCE SETS AND FUNGIBLE ESTIMATES}
        \fancyfoot{}}
    \fancypagestyle{plain}{
        \fancyhead[L]{}
        \fancyhead[R]{\thepage}
        \fancyfoot{}}
\begin{document}

%
\thispagestyle{title}

\textit{This version may have minor deviations from the final published version on Psychological Methods.}

\begin{center}
\vspace*{.5in}
   Parameter uncertainty in structural equation models:\\
    Confidence sets and fungible estimates\\
\vspace*{1in}
      Jolynn Pek\\
      The Ohio State University\\
      Hao Wu\\
      Boston College\\
\end{center}
\vspace*{1in}
\noindent   Jolynn Pek, Department of Psychology, The Ohio State University; Hao Wu, Department of Psychology, Boston College\\
\vspace*{12pt}
\indent
   Author note: Both authors contributed equally to this work and correspondence concerning this article should be addressed to Jolynn Pek, Department of Psychology, The Ohio State University, 1835 Neil Avenue, Columbus, OH 43210-1222. E-mail: pek.5@osu.edu. 
   The writing and refinement of this work was supported by the Natural Sciences and Engineering Research Council of Canada (NSERC) Discovery Grant RGPIN-04301-2014 and the Ontario Ministry of Research and Innovation Early Researcher Award ER15-11-004 awarded to Jolynn Pek.

\thispagestyle{fancy}
\setcounter{page}{2}
\begin{centering}
    \textbf{Abstract}\\
\end{centering}
    Current concerns regarding the dependability of psychological findings call for methodological developments to provide additional evidence in support of scientific conclusions. This paper highlights the value and importance of two distinct kinds of parameter uncertainty which are quantified by confidence sets (CSs) and fungible parameter estimates (FPEs; \citeNP{lee.maccallum&browne.inpress}); both provide essential information regarding the defensibility of scientific findings. Using the structural equation model, we introduce a general perturbation framework based on the likelihood function that unifies CSs and FPEs and sheds new light on the conceptual distinctions between them. \blue{A targeted illustration is then presented to demonstrate the factors which differentially influence CSs and FPEs, further highlighting their theoretical differences.} With three empirical examples on initiating a conversation with a stranger \cite{bagozzi&warshaw.1988}, posttraumatic growth of caregivers in the context of pediatric palliative care \cite{cadell.et.al.2014}, and the direct and indirect effects of spirituality on thriving among youth (\citeNP{dowling.et.al.2004}), we illustrate how CSs and FPEs provide unique information which lead to better informed scientific conclusions. Finally, we discuss the importance of considering information afforded by CSs and FPEs in strengthening the basis of interpreting statistical results in substantive research, conclude with future research directions, and provide example OpenMx code for the computation of CSs and FPEs.
    \\
    \vspace{12pt}
\newpage
\begin{centering}
Parameter uncertainty in structural equation models:\\
Confidence sets and fungible estimates\\
\end{centering}
\vspace{12pt}
   Statistical practice in psychological science is undergoing reform in response to concerns over the dependability of findings \cite{harlow.mulaik&steiger.2016, osc.2015, pashler&wagenmakers.2012,simmons.nelson&sinomsohn.2011,sijtsma.2015}. In response to these concerns, the reporting of effect sizes or focal parameter estimates and their confidence intervals (CIs) have been recommended as best practice \cite{cumming.2014, wilkinson.1999}. Confidence intervals communicate precision in estimation, providing information to researchers on whether inferences about their commensurate parameters can be drawn. Note that confidence regions (CRs) are an extension of CIs from a single parameter to a set of multiple parameters, and we use the term confidence sets (CSs) to collectively refer to CIs and CRs. Additionally, recent quantitative developments show that examining parameter sensitivity \cite{lee&maccallum.2015} and \emph{fungible} parameter estimates (FPEs; \citeNP{lee.maccallum&browne.inpress}; \citeNP{maccallum.browne&lee.2009}; \citeNP{waller.2008}) can add to substantive researchers' diagnostic toolkit in terms of generating more information about the \blue{validity} of their statistical results. Relative to CSs, FPEs are unfamiliar to substantive researchers; and although CSs and FPEs are different types of parameter uncertainty, little is known about FPEs in relation to CSs and how FPEs can be computed in practice.
   \\
   This paper has several aims. First, we provide a nontechnical overview of FPEs, their interpretation, and guidance regarding their computation. To better understand FPEs, this overview introduces an alternative perturbation to define FPEs (cf., \citeNP{lee.maccallum&browne.inpress,maccallum.lee&browne.2013,maccallum.browne&lee.2009}). Second, we emphasize the value in quantifying parameter uncertainty with the construction and computation of CSs and FPEs for drawing strong conclusions in substantive research. Third, we clarify the relationship between CSs and FPEs, and emphasize the distinct information each type of parameter uncertainty quantifies in SEM. Fourth, we provide example OpenMx code to construct CSs and FPEs in practice.
\\
   In SEM, parameter estimates are typically interpreted when a model is considered to fit the data well. The fit of a model is related to the amount of model error  (\citeNP{browne&cudeck.1993}; \citeNP{maccallum.2003}; \citeNP{wu&browne.2015a,wu&browne.2015b}) addressed by estimates of model fit indices such as the root mean square error of approximation (RMSEA, denoted as $\varepsilon$; \citeNP{steiger.2016,browne&cudeck.1993}), the Comparative Fit Index (CFI; \citeNP{bentler.1990}), and the Tucker-Lewis index (TLI; \citeNP{tucker&lewis.1973}). All of these indicators of model fit stem from the estimated model discrepancy, $\hat{F}$. \blue{Note that the term model fit here represents a quantification of how distant the specified model, which is a collection of distributions indexed by parameters, is from the sample or population. It is usually taken to be the smallest discrepancy between the given sample or population and the model, and the set of parameter values which indexes the member of the model which gives this smallest discrepancy is considered the best estimates.} Given good model fit where \blue{the model serves as a good representation of the sample}, CSs should then be constructed so as to quantify sampling variability or estimate precision about the population parameters. Tighter CSs imply less sampling variability, higher estimate precision, and a stronger tendency to conclude statistical significance. Finally, as part of model diagnostics, \blue{the influence of parameters in relation to the model's fit to data} should be considered so that \blue{valid} conclusions about parameter estimates can be made with confidence \cite{green.1977}. Fungible parameter estimates (FPEs) quantify parameter \blue{influence on the model's fit to the data in terms of variability in the values parameters can take within the small neighborhood of the optimal estimate. These FPEs are associated with the same specified model and serve as  alternative parameter values which describe the data just slightly worse than the unique set of optimal parameter estimates which achieves the best fit. Tighter sets of FPEs indicate less variability or robustness in the specified model's description of the data afforded by the alternative parameter estimates, providing a stronger basis for the validity of the optimal parameter estimates.} Confidence sets and the set of FPEs are important and distinct expressions of parameter uncertainty to be quantified and evaluated when strong scientific conclusions are sought (e.g., see \citeNP{pek.chalmers&monette.2016} for developments in multiple linear regression).
\\
   We begin by reviewing the SEM to establish notation, emphasizing maximum likelihood (ML) estimation. Next, we review profile likelihood CSs and FPEs and introduce a general framework based on the likelihood function that unifies CSs and FPEs in SEM as different aspects of parameter uncertainty, and showing how likelihood-based CSs and FPEs are analytically related. \blue{We then make use of a targeted illustration to demonstrate how different modeling factors influence CSs and sets of FPEs, emphasizing the distinct information each quantification of parameter uncertainty carries.} Next, we illustrate the application of CSs and FPEs with three real data examples on initiating a conversation with a stranger \cite{bagozzi&warshaw.1988}, posttraumatric growth in caregivers of children with life-limiting illnesses \cite{cadell.et.al.2014}, and the direct and indirect effects of spirituality on thriving among youth (\citeNP{dowling.et.al.2004}). Finally, we discuss the importance of considering CSs and FPEs in interpreting results obtained from the application of SEM, and conclude with future research directions.
\\
\vspace{12pt}
\begin{centering}
   \textbf{The Structural Equation Model}\\
\end{centering}
\vspace{12pt}
  Structural equation models are multi-parameter models where a system of linear equations among sets of measured and latent variables (MVs and LVs) are specified. The $p \times p$ population covariance matrix of the MVs is denoted by $\bm{\Sigma}$, and the model impled $k \times 1$ vector of parameters is denoted by $\bm{\theta}$. The SEM is expressed as $\bm{\Sigma} = \bm{\Sigma}(\bm{\theta})$, implying that the population covariance matrix for the MVs is a function of the model parameters. Parameter estimates $\hat{\bm{\theta}}$ are computed by minimizing the discrepancy between the model impled population matrix $\bm{\Sigma}(\bm{\theta})$ and the sample covariance matrix $\bm{S}$, which is expressed as a discrepancy function, $F[\bm{\Sigma}, \bm{S}]$. \blue{Observe that this function relates parameter values to a quantification of fit to a given sample.}
\\
  Many discrepancy functions for computing parameter estimates have been devised such as generalized least squares (GLS) and asymptotic distribution free (ADF; \citeNP{browne.1984}). The method of maximum Wishart likelihood (MWL) continues to be the most popular approach, assuming that the $p$ MVs follow a multivariate normal distribution. Minimizing the MWL discrepancy function is equivalent to maximizing the likelihood function of the sample covariance matrix $\bm{S}$ (see \citeNP{bollen.1989}, p.134--135). The MWL discrepancy function is
\begin{equation}\label{eq:1}
     F[\bm{\Sigma}, \bm{S}] = \ln|\bm{S}| + tr(\bm{S}\bm{\Sigma}^{-1}) - \ln|\bm{\Sigma}| - p,
\end{equation}
   where $p$ is the number of MVs and $tr(\cdot)$ denotes the trace of a square matrix.
\\
   The estimated sample discrepancy function is denoted as $\hat{F} = F[\bm{\Sigma(\hat{\theta})},\bm{S}]$ and multiplying $\hat{F}$ by $(N-1)$, where $N$ is the sample size, obtains the goodness-of-fit test statistic:
    \begin{equation*}
       X^2 = (N-1)\hat{F}.
    \end{equation*}
   This $X^2$ test statistic asymptotically follows a $\chi^2$ distribution with $p(p+1)/2-k$ degrees of freedom under the null hypothesis that the model implied covariance matrix $\bm{\Sigma}(\bm{\theta})$ is exactly the same as the population covariance matrix $\bm{\Sigma}$, $H_0: \bm{\Sigma} = \bm{\Sigma}(\bm{\theta})$. This goodness-of-fit test statistic, which quantifies a model's fit to the data, is also a likelihood ratio test (LRT) statistic because $X^2$ is twice the difference between two log-likelihoods (see Equation \ref{eq:1}). 
   Inverting a similar LRT statistic which compares two nested models is one basis of quantifying parameter uncertainty from the likelihood function.
\\
\vspace{12pt}
\begin{centering}
   \textbf{Quantifying Parameter Uncertainty}\\
\end{centering}
\vspace{12pt}
   \noindent\textbf{Confidence Sets}\\
   Recall that confidence sets collectively refer to CIs for a single parameter and CRs for multiple parameters, and supplement point estimates by conveying their estimate precision and sampling variability. \blue{Although CSs are estimated from sample data, they are devices for statistical inference in that they are statements about unknown population parameters.} 
   Formally, a CS provides a range of plausible population values for the set of parameters $\bm{\theta}$, and $(1-\alpha)100\%$ of such CSs are expected to contain the vector of population parameters over repeated sampling.
   There are several approaches to construct CSs and we focus on the profile likelihood method.
   \\
   \textbf{Profile likelihood confidence intervals}. Typically, SEMs have nuisance parameters which are required to complete the model and are not of substantive interest (e.g., unique variances in a factor analytic model). For simplicity, suppose that the vector of model parameters $\bm{\theta}$ is partitioned into two sets: a single focal parameter $\theta_f$, and a set of nuisance parameters $\bm{\theta}_n$. A profile likelihood CI for a $k_f=1$ focal parameter is constructed by inverting a LRT which tests the null hypothesis $H_0: \theta_f = \theta_{0f}$, where $\theta_{0f}$ is a scalar value of the population parameter defined under the null hypothesis. The LRT statistic is defined as $G^2 = 2[l(\hat{\bm{\theta}})-l(\theta_{0f},\tilde{\bm{\theta}}_n)]$,
   where $l(\hat{\bm{\theta}})$ is the log-likelihood associated with the $k \times 1 $ vector of ML estimates $\hat{\bm{\theta}}=(\hat{\theta}_f, \hat{\bm{\theta}}_n)'$, and $l(\theta_{0f},\tilde{\bm{\theta}}_n)$ is the log-likelihood associated with the ML estimates $(\theta_{0f},\tilde{\bm{\theta}}_n)'$, where $\theta_{0f}$ is held fixed  under the null hypothesis. Note that the vector of estimates for the nuisance parameters $\hat{\bm{\theta}}_n$ is distinct from $\tilde{\bm{\theta}}_n$ because $\hat{\bm{\theta}}_n$ is estimated jointly with $\hat{\theta}_f$ whereas $\tilde{\bm{\theta}}_n$ is estimated while $\theta_f=\theta_{0f}$ is held fixed.
   Given the observed data, a profile likelihood is an expression of values of $G^2$ in relation to the focal parameter $\theta_{0f}$, where the nuisance parameters $\bm{\theta}_n$ have been eliminated; $\bm{\theta}_n$ is eliminated by computing their ML estimates, $\tilde{\bm{\theta}}_n$, for each fixed value of $\theta_{0f}$ which relates to a $G^2$ value.
\\
    \blue{Under the assumption of multivariate normality,} this LRT statistic, $G^2$, asymptotically follows a $\chi^2$ distribution with $k_f=1$ degree of freedom. The profile likelihood CI is an inversion of $G^2$ because the CI is the set of values of $\theta_{0f}$ that satisfy $ G^2 \leq \chi^2_{1,1-\alpha}$. The right side of this inequality is the critical value based on the $\chi^2$ distribution
    where the first subscript denotes the degrees of freedom, and the second subscript ($1-\alpha$) denotes the confidence level. The unique upper and lower bounds of the CI are defined when $G^2 = \chi^2_{1,1-\alpha}$. In practice, because the profile likelihood for $\theta_f$ is obtained numerically, due to the re-estimation of $\tilde{\bm{\theta}}_n$ for different fixed values of $\theta_{0f}$, the search for values of $\theta_{0f}$ such that $G^2 = \chi^2_{1,1-\alpha}$ is also a numerical one. A faster search algorithm, which does not involve re-estimating $\tilde{\bm{\theta}}_n$, was proposed by Wu and Neale \citeyear{wu&neale.2012} and reviewed by Pek and Wu \citeyear{pek&wu.2015}.
\\
     \textbf{Profile likelihood confidence regions}. Constructing a profile likelihood CR for $k_f >1$ parameters is a direct extension of the $k_f=1$ case. The $G^2$ LRT statistic is modified into a joint test where the null hypothesis, $H_0: \bm{\theta}_{f} = \bm{\theta}_{0f}$, is an expression of vectors and
\begin{equation}\label{eq:2}
    G^2 = 2[l(\hat{\bm{\theta}})-l(\bm{\theta}_{0f},\tilde{\bm{\theta}}_n)].
\end{equation}
    Accordingly, a $(1-\alpha)100\%$ CR for $k_f>1$ focal parameters is defined as the set of values of $\bm{\theta}_{0f}$ that satisfy the inequality
\begin{equation}\label{eq:3}
    G^2 \leq \chi^2_{k_f, 1-\alpha}.
\end{equation}
    Observe that the degrees of freedom associated with the critical value based on the $\chi^2$ distribution are now $k_f$ (cf., \citeNP{scheffe.1953}). Similarly, boundary points to the CR are uniquely defined when $G^2 = \chi^2_{k_f, 1-\alpha}$.
\\
    \textbf{Computing profile likelihood confidence sets}. 
    In practice, computing a high dimensional CS becomes computationally untenable as the number of dimensions in $k_f$ increase, and presenting and interpreting such results become complex. Researchers have typically addressed the challenge of high-dimensionality by focusing on a limited number of focal parameters (e.g., \citeNP{lee.maccallum&browne.inpress}) or by computing $k_f$ simultaneous CIs which are one-dimensional projections of the CS (e.g., \citeNP{pek.chalmers&monette.2016}; \citeNP{pek&wu.2015}). We primarily focus on the method of a limited number of focal parameters.
    \\
\vspace{12pt}
   \noindent \textbf{Fungible Parameter Estimates}\\
   Fungible parameter estimates provide diagnostic information about statistical results from the broader framework of sensitivity analysis (\citeNP{cook.1986}; \citeNP{lee&wang.1996}). \blue{Specifically, FPEs inform of whether their commensurate parameter estimates should be validly interpreted.} In a sensitivity analysis, one introduces small perturbations to the data (e.g., deleting a case to quantify case influence; \citeNP{pek&maccallum.2011}) or to the model (e.g., perturbing parameter estimates to quantify parameter sensitivity; \citeNP{cadigan.1995}; \citeNP{lee&maccallum.2015}) and examine the effects of such perturbations on statistical results. The goal of sensitivity analysis is to determine the extent to which results are sensitive or robust to such small perturbations. Stability of statistical results are sought under controlled perturbations to the data or model \blue{ so as to establish their validity}. Conversely, observed instability of results under small introduced perturbations can undermine confidence in their definitive interpretation.
\\
    In a seminal paper, Green \citeyear{green.1977} stated that parameters with large effects on the fit of a \blue{single specified} model provide a strong basis for scientific conclusions. \blue{The rationale behind this statement stems from the link between values a model's parameters can assume (i.e., $\hat{\bm{\theta}})$ and their implied fit to the data (e.g., $\hat{F}$). Parameters closely tied to model fit play an essential role in determining how well the model describes the data; their strong influence on the specified model's fit to data serves to validate their rigorous interpretation. Conversely, parameters which do not exert influence on model fit have little basis for interpretation because they are relatively uninvolved with how well the model represents the data \cite{lee&maccallum.2015}. This influence or sensitivity of parameters can be determined} from a sensitivity analysis where optimal focal parameter estimates $\hat{\bm{\theta}}_f$ are perturbed to quantify their influence on model fit (\citeNP{lee&wang.1996}); parameters which strongly influence model fit are deemed sensitive whereas the converse holds for insensitive parameters. 
\\
    Fungible parameter estimates approach the issue of parameter sensitivity by perturbing model fit instead of optimal parameter estimates. \blue{Thus, given a slight and fixed perturbation to a specified model's fit to data}, FPEs are alternative parameter estimates \blue{located within the small neighborhood of the optimal estimate that yield the same level of perturbed model fit. Stated differently, the FPE approach considers a discrepancy function value slightly greater than the minimum and its associated parameter values; the suboptimal discrepancy function value is considered an alternative model fit, and the commensurate parameter values (i.e., FPEs) are alternative estimates.} Tight sets of FPEs with limited variability reflect influential parameters. Conversely, highly varying sets of FPEs reflect insensitive parameters.
    \blue{The key diagnostic information communicated by FPEs is the stability of the model's description of the data afforded by the parameter estimates $\hat{\bm{\theta}}$ in relation to estimated model fit. Here, stability or robustness enhances the validity of parameter estimates.} 
\\
    When a specified model is fit to sample data, optimal parameter estimates $\hat{\bm{\theta}}$ are obtained at the minimum value of the discrepancy function, $\hat{F}$. Typically, model fit is assessed by making a judgment on the value of $\hat{F}$ or some function thereof, and parameter estimates $\hat{\bm{\theta}}$ are interpreted when the model is deemed to fit the data well. 
  Fungible parameter estimates are regarded as alternative estimates which yield the same value of suboptimal model fit, $\hat{F}^*$; because this is a type of sensitivity analysis, suboptimal model fit is broadly defined as slightly worse than optimal model fit where the difference between optimal and suboptimal fit is of no substantive consequence (\citeNP{lee.maccallum&browne.inpress}). The variability of the set of FPEs from the ML estimate $\hat{\bm{\theta}}$ quantifies the stability of the model's description of the data, in terms of parameter estimates, under a slight perturbation to model fit. Wildly varying FPEs imply that parameter estimates can take on very different values (cf., insensitive parameters) and consequently very different interpretations, \blue{raising questions about their validity and reducing confidence in drawing inferences about them}. Conversely, invariable FPEs (cf. sensitive parameters) suggest that parameter estimates are stable and remain relatively consistent under a slight perturbation to model fit, \blue{establishing their validity and} strengthening confidence in their rigorous interpretation.\blue{\footnote{\blue{From the perspective of estimation where a function relating model fit to parameter values is optimized (e.g., Equation 1), A set of FPEs contain alternative parameter values associated with a suboptimal solution. Suppose the likelihood is to be optimized such that unique optimal parameter estimates are ML estimates. By perturbing the maximum likelihood value which is also a quantification of model fit (e.g., minus two the log-likelihood, $-2LL$), FPEs are a `slice' of the likelihood and communicate the extent of information present in the data about the parameters (see \citeNP{myung.2003} for a tutorial on ML estimation). Highly varying FPEs communicate limited information whereas invariant FPEs communicate high information, and more information augments the validity of the commensurate parameter estimates.}}} Note that the interpretation of parameter estimates requires alignment with substantive theory and related conceptual and philosophical issues \cite{bagozzi&yi.1988}. To the extent that FPEs corroborate substantive theory and concepts as alternative parameter estimates, they provide evidence on the stability of the model's description of the data.
\\
   Let $\hat{\bm{\theta}}^*$ denote the set of FPEs which are uniquely defined when a slight perturbation is added to the value of the estimated discrepancy function value $\hat{F}$, denoted as $\hat{F}^* = \hat{F}+\delta_{FPE}$, where $\delta_{FPE}$ is a non-random and fixed value. Similar to the arbitrary choice of 1\%, 5\% or 10\% significance levels, the magnitude of the perturbation $\delta_{FPE}$ to apply is subjective but should be consistent with the notion that perturbations are necessarily small to the extent that suboptimal model fit is practically no different from the fit of the optimal solution. \blue{We are reluctant to ascribe guidelines for recommended magnitudes of $\delta_{FPE}$ because the value of constructing sets of FPEs lies in the extent of stability or uncertainty communicated by their variability associated with a range of levels of perturbation.} Indeed, the construction of FPEs is for the diagnostic purpose of ascertaining \blue{whether parameters can be validly interpreted} based on the extent of parameter sensitivity or the stability of the parameter estimates under a slight perturbation to model fit. In this vein, several levels of perturbation are recommended for practice \cite{lee.maccallum&browne.inpress}. Here, we focus on two types of perturbations with different properties: the first is based on the RMSEA, and the second is based on $\hat{F}$.
\\
   Extant research on FPEs forward that because a perturbation to model fit is more interpretable in the scale of the RMSEA compared to $\hat{F}$ \cite{lee.maccallum&browne.inpress, maccallum.browne&lee.2009}, the perturbation $\delta_{FPE}$ can be defined in the metric of the RMSEA. The RMSEA is a measure of discrepancy per degree of freedom and takes into account model complexity \cite{browne&cudeck.1992,steiger.2016}. The population RMSEA is $\epsilon = \sqrt{F_0/df}$, where $F_0$ is the discrepancy due to model error in the population and $df = p(p+1)/2 - k$ is the model degrees of freedom. The sample RMSEA, which corrects for the bias in $\hat{F}$ as an estimator of $F_0$, is
 $
        \hat{\epsilon} = \sqrt{\mbox{max}(\frac{\hat{F}}{df}-\frac{1}{N-1},0)}
  $
  where max($\cdot$) is the maximum operator \cite{browne&cudeck.1992}. The range $ \hat{\epsilon} \leq$ 0.05 is conventionally viewed as indicating a close fit of the model to the data relative to the model degrees freedom. Note that when $F_0$ is close to zero, and sample size is small, $\hat{\epsilon}$ is truncated at zero because the bias correction term $\frac{1}{(N-1)}$ could be larger than $\frac{\hat{F}}{df}$.
  \\
    Let the perturbed value of RMSEA be $\epsilon^* = \hat{\epsilon} + \tilde{\epsilon}$, where $\tilde{\epsilon}$ denotes the perturbation to the optimal RMSEA. Several levels of perturbation are recommended in a sensitivity analysis; previous research has used $\tilde{\epsilon} = .001$ and .005 based on the rationale that such perturbations do not substantively change model fit (e.g., see \citeNP{lee.maccallum&browne.inpress}, \citeNP{maccallum.lee&browne.2013}; cf. \citeNP{lee&maccallum.2015}). For example, $\hat{\epsilon} = .05$, $\hat{\epsilon} = .051$, and $\hat{\epsilon} = .055$ suggest inconsequential differences in model fit as operationalized by the RMSEA. The perturbed discrepancy function value is then
        \begin{equation}\label{eq:4}
            \hat{F}^* = df[(\hat{\epsilon}+\tilde{\epsilon})^2 + \frac{1}{N-1}],
        \end{equation}
    when the estimated RMSEA $\neq 0$. Although FPEs should theoretically reflect properties of the model and be independent of sample size (cf., \citeNP{waller.2008}), the perturbation based on the RMSEA in the scale of $F$, denoted as $\delta_\epsilon = \hat{F}^*-\hat{F}$, is affected by sample size $N$ because of the bias correction. This dependence will  be discussed later.\footnote{For a perturbation based on the RMSEA to be free of sample size, the RMSEA without the bias correction can be perturbed instead. Note that Waller's \citeyear{waller.2008} fungible weights in multiple linear regression are independent of sample size because the perturbation is based on the unadjusted $R^2$; if the bias adjusted $R^2$ is perturbed, resulting fungible weights will also be affected by sample size.}  
     Note that FPEs are formulated to provide diagnostic information about the stability of a particular model's parameter estimates, and the property of invariance for the perturbation $\delta_{FPE}$ across samples and models is unimportant in this context.
\\
    An alternative perturbation to define FPEs, which is explicitly free from sample size and $df$, can be applied directly to the sample discrepancy function value $\hat{F}$. Specifically, a small percentage of $\hat{F}$ can be used as a perturbation. Under this alternative scheme the perturbed discrepancy function value is
        \begin{equation}\label{eq:5}
                \hat{F}^* = \hat{F} + \delta_F,
        \end{equation}
        where $\delta_F$ is some small percentage of $\hat{F}$. For instance, when the perturbation is defined by 5\% or $\delta_F = .05\hat{F}$, $\hat{F}^* = 1.05\hat{F}$. Unlike $\delta_\epsilon$, $\delta_F$ is free of sample size (for the same observed covariance matrix) and its resulting FPEs are free from the influence of sample size and purely reflect stability of the parameter estimates. In contrast to defining a perturbation in the scale of RMSEA, $\tilde{\epsilon}$, the scale of $F$ tends to be more unfamiliar to researchers although it can be considered a sum of squared standardized residuals under certain regularity conditions \cite{steiger.2016}. The magnitude of $\delta_F$, which is a percentage of $\hat{F}$, directly depends on the size of $\hat{F}$ such that larger $\hat{F}$ would be associated with larger $\delta_F$ in magnitude, and vice-versa. 
        Consistent with any sensitivity analysis, several levels of $\delta_F$ (e.g., $.02\hat{F}$ and $.05\hat{F}$) should be used in practice to adequately gauge the stability of focal parameter estimates under slight perturbations to model fit.
        \\
    Regardless of the form of the perturbation $\delta_{FPE}$, which can be $\delta_\epsilon$ or $\delta_F$, FPEs for the $k$ parameters, denoted by the vector $\hat{\bm{\theta}}^*=\hat{\bm{\theta}}+\bm{\varepsilon}$, are obtained by solving
        \begin{equation*}
            F[\bm{\Sigma}(\hat{\bm{\theta}}^*, \bm{S})] = \hat{F}^*,
        \end{equation*}
    or equivalently,
        \begin{equation}\label{eq:6}
            F[\bm{\Sigma}(\hat{\bm{\theta}}+\bm{\varepsilon}, \bm{S})] = \hat{F}+\delta_{FPE},
        \end{equation}
    where $\bm{S}$, $\hat{F}^*$ and $\hat{\bm{\theta}}$ are known, and the set of FPEs is obtained by perturbing the ML estimates $\hat{\bm{\theta}}$ by a set of vectors $\bm{\varepsilon}$ such that $\hat{F}^*$ is obtained. Each unique vector $\bm{\varepsilon}$ defines the magnitude and direction of where each unique FPE lies from $\hat{\bm{\theta}}$. In practice, the search for $\bm{\varepsilon}$ is a numerical one because no closed-form solution exists in SEM, and we highlight the profile likelihood approach below in the context of focal parameters (for a closed form solution to multiple linear regression, see \citeNP{pek.chalmers&monette.2016}). 
 \\
   \textbf{A pair of profile likelihood fungible estimates}. Consider computing a pair of FPEs about a single focal parameter $\theta_f$; this pair of scalar FPEs will be defined by two unique and distinct values of $\varepsilon$ that fall below ($\varepsilon_L$) and above ($\varepsilon_U$) the optimal estimate $\hat{\theta}_f$. With the partitioning of focal and nuisance parameters $\bm{\theta} = (\theta_f,\bm{\theta}_n)'$, the definition of FPEs in Equation \ref{eq:6} is modified to reflect this partitioning. Specifically, the lower FPE value is defined as $ F[\bm{\Sigma}(\hat{\theta}_f+\varepsilon_L, \tilde{\bm{\theta}}_n),\bm{S}] = \hat{F}+\delta_{FPE}$. Similarly, the upper FPE value is defined by substituting $\varepsilon_L$ with $\varepsilon_U$. Note that $\varepsilon_L$ is necessarily negative and $\varepsilon_U$ is necessarily positive. By reducing the dimensionality of $k$ parameters to a $k_f=1$ focal parameter, the search for FPEs is reduced to a single dimension in the parameter space. 
   \\
   \textbf{A profile likelihood fungible contour}. Typically, researchers are interested in a limited number of focal parameters in SEM. Computing a set of FPEs which form a contour for $k_f >1$ parameters is a direct extension of the $k_f=1$ case. In particular, the scalar $\theta_f$ is generalized to a vector $\bm{\theta}_f$ of length $k_f$; likewise, the two scalar values of $\varepsilon_L$ and $\varepsilon_U$ are generalized to a set of multiple vectors, each denoted as $\bm{\varepsilon}$ with length $k_f$, that span the $k_f$-dimensional space. Formally, FPEs for $k_f$ parameters is expressed as
    \begin{equation}\label{eq:7}
        F[\bm{\Sigma}(\hat{\bm{\theta}}_f+\bm{\varepsilon}, \tilde{\bm{\theta}}_n),\bm{S}] = \hat{F}+\delta_{FPE}.
    \end{equation}
   The vector $\bm{\varepsilon}$ quantifies the magnitude and direction with which $\hat{\bm{\theta}}_f$ is perturbed to satisfy the slightly perturbed model fit value $\hat{F}^*$.
 \\
   \textbf{Computing profile likelihood fungible estimates}.
    \citeA{maccallum.lee&browne.2013} introduced a root finding algorithm by Brent (\citeyearNP{brent.1973}; cited in \citeNP{press.et.al.1992}) as one approach to compute FPEs\footnote{MacCallum et al.'s \citeyear{maccallum.lee&browne.2013} computational method is distinct from the profile likelihood method outlined in this paper in that the nuisance parameters $\bm{\theta}_n$ are held fixed at $\hat{\bm{\theta}}_n$ instead of re-estimated as $\tilde{\bm{\theta}}_n$. Their approach likely results in the conservative approximation of the size of fungible contours because the uncertainty of the nuisance parameters is not taken into account \cite{lee.maccallum&browne.inpress}.}. This root finding algorithm was adapted from previous work by MacCallum, Lee, and Browne \citeyear{maccallum.lee&browne.2010} on the phenomenon of isopower in SEM, and has also been included in a review of likelihood-based methods to compute CSs (see Pek \& Wu, \citeyearNP{pek&wu.2015}). Alternatively, FPEs can be estimated via the faster algorithm developed by Wu and Neale \citeyear{wu&neale.2012}, which does not involve the nested iterative estimation of $\tilde{\bm{\theta}}_n$.  This algorithm has been implemented in the current version of OpenMx \cite{neale.et.al.2015} for profile likelihood CIs. Indeed, the estimation of profile likelihood CSs and FPEs can make use of the same algorithms because both types of parameter uncertainty can be unified under a general perturbation framework based on the likelihood function (see also supplemental material).
 \\
 \vspace{12pt}
   \begin{centering}
   \textbf{A General Perturbation Framework}\\
   \end{centering}
   \vspace{12pt}
   Parameter uncertainty is fully represented by the likelihood function \cite{pawitan.2001}, such that profile likelihood CSs and FPEs can be unified under a general framework (cf., \citeNP{pek.chalmers&monette.2016}). First, observe that parameters $\bm{\theta}$ can be represented as a function of some form of the likelihood such as $F$ from Equation \ref{eq:1} (see Bollen, \citeyearNP{bollen.1989}, p. 131--135 for the derivation of the discrepancy function from the log-likelihood). Next, the computation of a boundary point of a CS or an FPE for focal parameters $\bm{\theta}_f$ can be couched as a perturbation from the optimal focal estimates $\hat{\bm{\theta}}_f$ by the vector $\bm{\varepsilon}$ due to a perturbation $\delta$ applied to the likelihood value associated with the optimal model $\hat{F}$:
         \begin{equation}\label{eq:8}
            F[\bm{\Sigma}(\hat{\bm{\theta}}_f, +\bm{\varepsilon}, \tilde{\bm{\theta}}_n),\bm{S}] = \hat{F} + \delta,
         \end{equation}
    where the nuisance parameters are eliminated by their ML estimation in $\tilde{\bm{\theta}}_n$. Geometrically, the perturbation $\delta$ is the vertical distance from the maximum point of the likelihood, which defines a horizontal `slice' of the profile likelihood surface expressed in $F$.
 \\
 \vspace{12pt}
    \noindent\textbf{Confidence Sets}\\
  The definition of CSs for $k_f$ focal parameters from Equations \ref{eq:2} and \ref{eq:3} can be re-expressed to follow the form of Equation \ref{eq:8} of the general perturbation framework. Specifically, the LRT statistic $G^2$, which is inverted to construct CSs, can be re-expressed in the metric of the MWL discrepancy function: $G^2 = (N-1)[F[\bm{\Sigma}(\bm{\theta}_{0f}, \tilde{\bm{\theta}}_n), \bm{S}]-\hat{F}]$, where $\bm{\Sigma}(\bm{\theta}_{0f}, \tilde{\bm{\theta}}_n)$ is the population covariance matrix holding the focal parameters $\bm{\theta}_f$ fixed at values $\bm{\theta}_{0f}$ under the null, while the nuisance parameters are eliminated by re-estimation to obtain $\tilde{\bm{\theta}}_n$. Recall that $\bm{\theta}_{0f}$ are boundary points to the CS for the $k_f$ focal parameters, which can be expressed as perturbations from the optimal focal parameter estimates, $\bm{\theta}_{0f} = \hat{\bm{\theta}}_f + \bm{\varepsilon}$. With further algebraic manipulations, CSs can be alternatively defined as
    \begin{equation}\label{eq:9}
        F[\bm{\Sigma}(\hat{\bm{\theta}}_{f} + \bm{\varepsilon}, \tilde{\bm{\theta}}_n), \bm{S}]= \hat{F} + \chi^2_{1-\alpha,k_f}/(N-1).
    \end{equation}
    In relation to Equation \ref{eq:8}, the perturbation of $\delta$ to $\hat{F}$, which uniquely defines CSs, is the scaled critical value associated with $G^2$: $\delta_{CS}=\chi^2_{1-\alpha,k_f}/(N-1)$, where the subscript $CS$ denotes the perturbation that defines CSs. Because CSs are a special expression within the general perturbation framework of Equation \ref{eq:8}, boundary points to CSs can be taken as alternative parameter estimates that are perturbed by $\bm{\varepsilon}$ from optimal estimates $\hat{\bm{\theta}}_f$ due to a perturbation of optimal model fit $\hat{F}$ by a magnitude of $\chi^2_{1-\alpha,k_f}/(N-1)$. The perturbation defining CSs is a function of the critical value associated with the $G^2$ test statistic \blue{that requires the assumption of multivariate normality}. Note that this perturbation is inversely related to sample size: a larger sample size produces a smaller perturbation and therefore a smaller CS.
 \\
 \vspace{12pt}
   \noindent\textbf{Fungible Parameter Estimates}\\
   Similar to CSs, the expression of FPEs also follow the form of Equation \ref{eq:8} of the general perturbation framework. When a perturbation is added to the RMSEA to define FPEs (see Equation \ref{eq:4}), the perturbation in Equations \ref{eq:7} and \ref{eq:8} is $\delta = \delta_{\epsilon}$, i.e.,
 \begin{equation}\label{eq: delta_e}
   \delta_\epsilon = df \left\{(\hat{\epsilon}+\tilde{\epsilon})^2+(N-1)^{-1}\right\}-\hat{F}
   		 = \left\{ \begin{array}{lc}
   		 df\tilde{\epsilon}^2+(N-1)^{-1}df-\hat{F} &	\mbox{if }\hat{\epsilon}=0\\
   		 df\left(\tilde{\epsilon}^2+2 \hat{\epsilon} \tilde{\epsilon}\right) & \mbox{if }\hat{\epsilon}>0
   		 \end{array} \right.
  \end{equation}
    Equation~\ref{eq: delta_e} sheds light on the way sample size $N$ may affect the size of sets of FPEs. We note that for $\hat{\epsilon}>0$, this perturbation does not explicitly involve sample size, but it is ultimately affected by sample size because $\hat{\epsilon}$ is involved in the linear term. For a fixed sample covariance matrix $\bm{S}$, when $\hat{\epsilon}=0$, $\delta_\epsilon$ decreases with increasing sample size until it reaches its minimum value of $df\tilde{\epsilon}$; when $\hat{\epsilon}$ becomes positive, $\delta_{\epsilon}$ increases with sample size because $\hat{\epsilon}$ increases with sample size. For samples of increasing sizes from a given population, this perturbation varies due to sampling error around a constant value when $\hat{\epsilon}$ is mostly positive but has a random but decreasing trend when $\hat{\epsilon}$ has a sizable chance of zero.
\\
   When FPEs are defined by a perturbation as a percentage of $\hat{F}$ (see Equation \ref{eq:5}), the perturbation in Equations \ref{eq:7} and \ref{eq:8} is $\delta = \delta_F$. For a fixed sample covariance matrix $\bm{S}$, this perturbation is constant and does not change with sample size or $df$. For samples of increasing sizes from a given population, this perturbation is random with a decreasing trend.

   The two definitions of FPEs, based on $\delta_\epsilon$ and $\delta_F$, are special expressions within the general perturbation framework of Equation \ref{eq:8}. \blue{In contrast to the perturbation defining CSs, perturbations defining FPEs are free from distributional assumptions because $\delta_{FPE}$ is not defined through any sampling distribution.} Similar to CSs, however, FPEs can be interpreted as perturbed values of $\bm{\varepsilon}$ from the optimal focal parameters $\hat{\bm{\theta}}_f$ due to a perturbation of magnitude $\delta_{\epsilon}$ or $\delta_F$ to optimal model fit, $\hat{F}$ or $\hat{\epsilon}$, respectively.
\\
   It is interesting to note the following relationship between $\delta_\epsilon$ and $\delta_F$.
    From Equation~\ref{eq: delta_e}, for $N=\infty$,
     \begin{equation} \label{eq:e_vs_f}
     \frac{\delta_\epsilon}{\hat F} = \frac{df\left(\tilde{\epsilon}^2+2 \hat{\epsilon}\tilde{\epsilon}\right)}{df\hat{\epsilon}^2}
     = \left(\frac{\tilde{\epsilon}}{\hat{\epsilon}}\right)^2+ 2\left(\frac{\tilde{\epsilon}}{\hat{\epsilon}}\right).
     \end{equation}
   This suggests that if we perturb RMSEA with no bias-adjustment proportional to its observed value, the result is equivalent to proportionally perturbing the observed discrepancy function value. For example, increasing the RMSEA (without bias adjustment) by $1\%$ would be equivalent to increasing $\hat{F}$ by about $2\%$, irrespective of the $df$ of the model or the value of RMSEA or $\hat{F}$.
 \\
 \vspace{12pt}
    \noindent\textbf{Analytical Relationship}\\
    Given that CSs and FPEs can be expressed as a function of the likelihood in the form of Equation \ref{eq:8}, we now consider the issue of when these two kinds of parameter uncertainty are numerically equivalent. Specifically, by equating the perturbations $\delta_{CS}$ with $\delta_\epsilon$ or $\delta_F$, CSs and FPEs are numerically equivalent if and only if
        \begin{equation} \label{eq:10}
            \chi^2_{1-\alpha,k_f}/(N-1) = df[(\hat{\epsilon}+\tilde{\epsilon})^2 + \frac{1}{N-1}] - \hat{F}
        \end{equation}
        or
               \begin{equation} \label{eq:11}
            \chi^2_{1-\alpha,k_f}/(N-1) = \delta_F .
        \end{equation}
    Although Equations \ref{eq:10} and \ref{eq:11} analytically establish the numerical equivalence of CSs and FPEs, these two kinds of uncertainty are \emph{neither} substantively \emph{nor} theoretically equivalent. \blue{Knowledge about one type of parameter uncertainty does not aid in the interpretation of the other type of uncertainty.} Equations \ref{eq:10} and \ref{eq:11} show that CSs and FPEs are different aspects of parameter uncertainty which can be quantified by the profile likelihood expressed in Equations \ref{eq:9} and \ref{eq:7}, respectively. More important, observe that the perturbation defining CSs is a function of the critical value of the LRT to be inverted and sample size; \blue{because LRTs are about population parameters, CSs are inferential devices}. In particular, sample size reflects sampling variability in that larger $N$ will reduce the magnitude of $\delta_{CS}$ and the size of the CS itself. In contrast, the magnitude of the perturbations which define FPEs ($\delta_\epsilon$ or $\delta_F$) are primarily arbitrarily determined by the analyst for the purpose of conducting a sensitivity analysis regarding the stability of the focal parameters; FPEs are diagnostic devices \blue{in that they provide information on whether optimal parameter estimates can be validly interpreted.}. The two definitions of FPEs ($\delta_\epsilon$ or $\delta_F$) measure model misfit differently and have contrasting properties: for a given sample covariance matrix, $\delta_\epsilon$ is affected by both model $df$ and sample size; in contrast, $\delta_F$ is free of $N$ and $df$, and is in a standardized scale of a percentage of $\hat{F}$.

\begin{landscape}
\begin{figure}[!htbp]
\centering
 \includegraphics[trim=0cm 0cm 0cm 0cm, clip=true, height=6in]{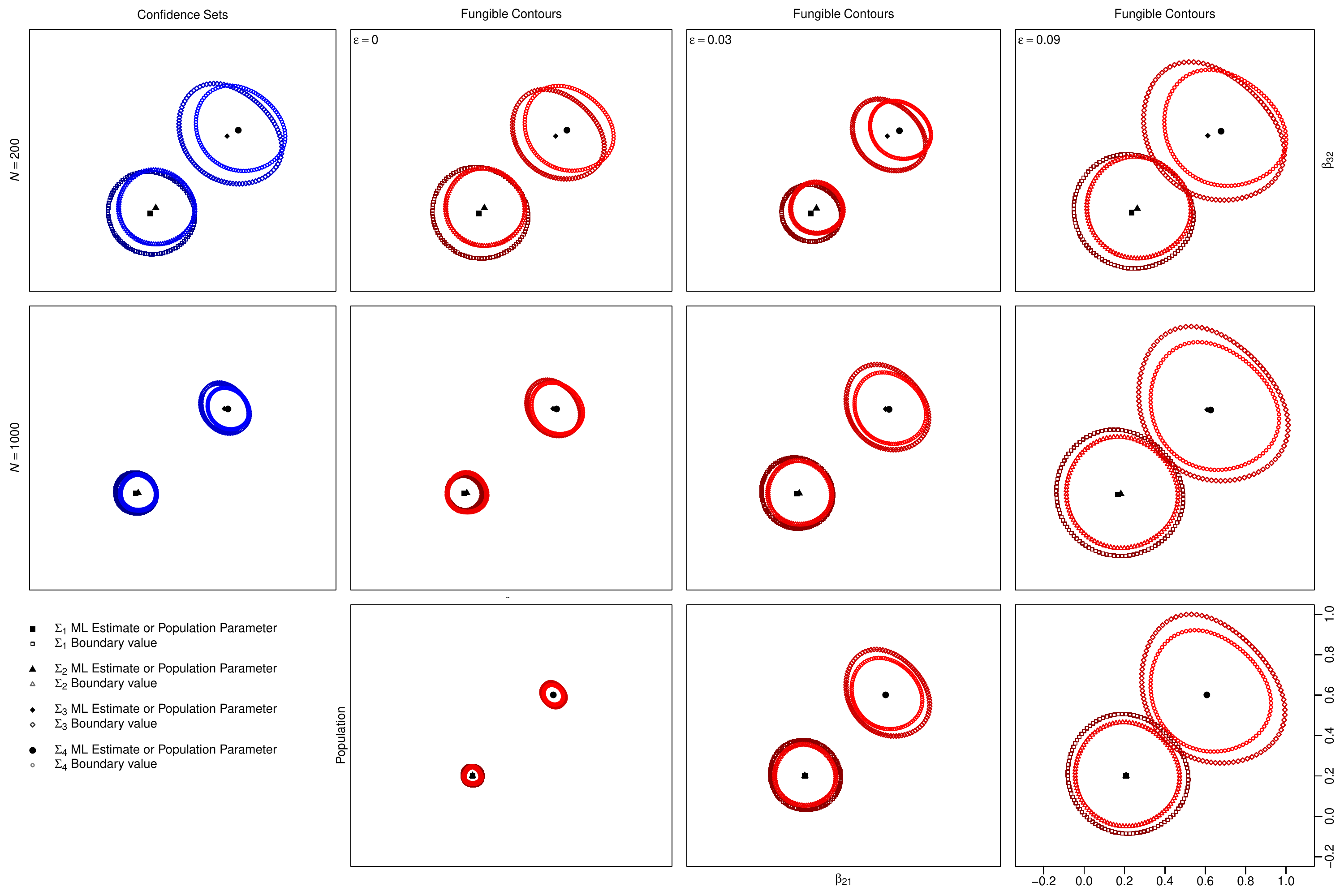}
 \caption{Profile likelihood confidence sets and fungible contours (defined by $\tilde{\epsilon}$) of $\beta_{21}$ and $\beta_{32}$ by sample size $N$ and type of model; $\epsilon$= 0 indicates perfect model fit, $\epsilon$=.03 indicates good model fit, and $\epsilon$=.09 indicates poor model fit; $\bm{\Sigma}_1$= large unique variances and small structural effects, $\bm{\Sigma}_2$= small unique variances and small structural effects, $\bm{\Sigma}_3$= large unique variances and large structural effects,
and $\bm{\Sigma}_4$= small unique variances and large structural effects.}
\label{fig:sim}
\end{figure}
\end{landscape}

\begin{figure}[!htbp]
\centering
 \includegraphics[trim=0cm 0cm 0cm 0cm, clip=true, height=7in]{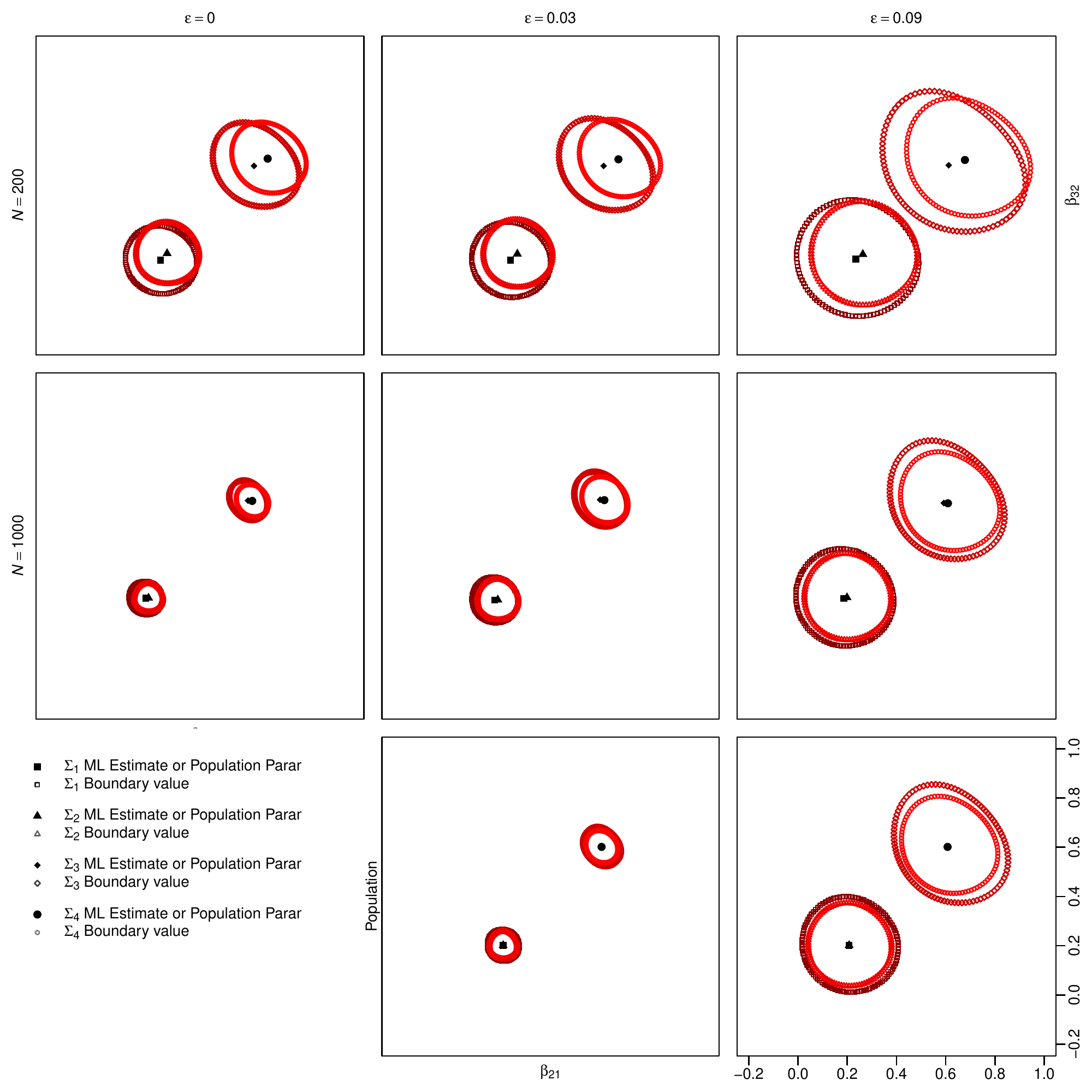}
 \caption{Fungible contours (defined by $\delta_F$ = $.05\hat{F}$) of $\beta_{21}$ and $\beta_{32}$ by sample size $N$ and type of model; $\epsilon$= 0 indicates perfect model fit, $\epsilon$=.03 indicates good model fit, and $\epsilon$=.09 indicates poor model fit; $\bm{\Sigma}_1$= large unique variances and small structural effects, $\bm{\Sigma}_2$= small unique variances and small structural effects, $\bm{\Sigma}_3$= large unique variances and large structural effects,
and $\bm{\Sigma}_4$= small unique variances and large structural effects.}
\label{fig:dF}
\end{figure}

\vspace{12pt}
  \noindent\textbf{Conceptual Distinctions}\\
    The two aspects of parameter uncertainty, quantified by CSs and FPEs, can be unified under a framework based on the likelihood function. Suppose that the direct effect of a latent predictor on a latent outcome is the focal parameter of interest in a LV mediation model (see supplemental material for details of this example based on \citeauthor{schmitt.et.al.2002}, 2002). After controlling for the latent mediator, the residual effect of the latent predictor on outcome is $-0.19$, 95\% $CI=[-0.39, -0.01]$. Although the CI implies a statistically significant and negative direct effect at the 5\% level of significance, the plausible population values for this parameter can be very close to zero. 
    Conversely, a perturbation of $\tilde{\epsilon} = .005$ or $\delta_F = .076\hat{F}$ in a sensitivity analysis is associated with FPE values of $-.533$ and 0.097. The FPEs suggest instability of the direct effect under a slight perturbation to model fit because the FPEs are widely varying, and the upper FPE suggests that the direct effect could be positive although the optimal estimate is negative. 
    In this instance, the FPEs suggest that the direct effect should not be \blue{validly interpreted and the CS implies high estimate imprecision about the population direct effect}. 
    \\
    Confidence sets allow researchers to make inferential statements about unknown population parameters because they span a range of plausible population values within their boundary points, which are limits to a range of estimates (see Equation \ref{eq:3}). Stated differently, the parameter values located on the boundary of the CS as well as within the CS are of substantive interest. Additionally, CSs communicate information about estimate precision and quantify sampling variability. 
    The information communicated by CSs informs researchers of how precise parameter estimates are, and CSs define the range of plausible values which the population parameters can take on.
\\
    In contrast, 
    FPEs are subjectively and arbitrarily defined, and there can be more than one definition of FPEs (e.g., $\tilde{\epsilon}$ versus $\delta_F$). Indeed, several sets of FPEs should be constructed to quantify the stability of a model's description of the data, afforded by the parameter estimates, with respect to a slight change in model fit. Note that the values which FPEs take on as alternative parameter estimates are not meant to be interpreted from a  substantive standpoint. Instead these FPEs are constructed to communicate the variability and stability in the values which parameter estimates can assume, which \blue{informs of whether optimal parameter estimates can be validly interpreted}. Several sets of FPEs provide important information on whether parameter estimates can be interpreted definitively \cite{lee.maccallum&browne.inpress}. When sets of FPEs with limited variability are obtained in a sensitivity analysis, strong conclusions can be drawn from the interpretation of significant focal parameters. When sets of FPEs with high variability are observed, which implies wildly different interpretations of effects that are associated with a slightly perturbed value of model fit, there is little basis to interpret the optimal estimates. Unlike CSs, where the region located within boundary points is meaningful as plausible population parameter values, only points in the set of FPEs are of interest; estimates lying within the contour or surface formed by the FPEs are associated with a perturbation to model fit that is smaller than what is specified. Important distinctions between CSs and FPEs are summarized in Table \ref{table:sum}.
    \\
    \vspace{12pt}
       \begin{centering}
    \textbf{Targeted Illustration}\\
    \end{centering}
    \vspace{12pt}
    \blue{To further emphasize the distinction between CSs and FPEs, we report on a targeted illustration} examining three factors which could influence the behavior of CSs and FPEs: (a) sample size, (b) model fit, and (c) the magnitude of correlations among the MVs. 
 \\
 \vspace{12pt}
    \noindent\textbf{Factors}\\
    \textbf{Sample size}. Sample size is expected to influence the size of CSs because the perturbation defining CSs decrease with increasing sample size. Increasing sample size would result in tighter CSs, reflecting higher precision and lower sampling variability. Sample size is expected to minimally influence the size of FPE sets as discussed. Two levels of sample size, $N=200$ and $N=1000$, were chosen to reflect moderate and large sample sizes typically observed in substantive research.
\\
    \textbf{Model fit}. Limited exploratory computations of FPEs on published examples have revealed a tendency for the size of FPE contours or surfaces to vary with model fit; larger FPE contours have been observed to be associated with less well-fitting models whereas tighter FPE contours have been observed to be associated with good fitting models. The estimated model discrepancy, $\hat{F}$, will be large under poor model fit, suggesting a flat likelihood surface. In this vein, decreasing levels of model fit is likely associated with decreasing peakedness of the likelihood surface. Alternatively, improving model fit is expected to be associated with tighter CSs and FPE contours. Three levels of model fit were examined, as defined by the population RMSEA: $\epsilon = 0$ (perfect fit), $\epsilon = 0.03$ (good fit), and $\epsilon = .09$ (poor fit). Overall model fit was controlled by using the method of Cudeck and Browne \citeyear{cudeck&browne.1992} 
    to construct a covariance matrix which yields a specified minimum discrepancy function value in the population, $F_0$, adding realism to the data generation process as actual data do not appear to follow models which hold exactly in the samples \cite{tucker.koopman&linn.1969}.
\\
    \textbf{Magnitude of correlations}. In general, larger correlations compared to smaller correlations among MVs are associated with more power to reject false models \cite{neale&miller.1997}. Fewer models can fit well to data structures with larger correlations among MVs, implying more peaked likelihood surfaces. Thus, it is hypothesized that data structures with larger correlations would result in tighter CSs and FPEs for focal parameters and vice-versa. Two levels of the magnitude of correlations among the MVs were specified (small versus large) and two approaches to control this magnitude were employed, resulting in four models. These two approaches are detailed in the section below.
\\
\vspace{12pt}
    \noindent\textbf{Population Generating Models}\\
    Data were generated based on a published LV mediation model by Schmitt et al. \citeyear{schmitt.et.al.2002}, where a LV mediation model was fit to 13 MVs, confirming that the effect of Perceived Discrimination on Well-being is mediated by In-group Identification\footnote{
    The supplemental material includes computations of CSs and FPEs for the direct and indirect effects of Perceived Discrimination on Well-being.}. The population generating model can be expressed as
        \begin{equation}
            \bm{\Sigma}=\bm{\Lambda}(\textbf{I}-\bm{B})\bm{\Phi}(\textbf{I}-\bm{B})' \bm{\Lambda}' + \bm{\Psi},
        \end{equation}
    where $\bm{\Lambda}$ is a $13\times3$ matrix of factor loadings, \textbf{I} is a $3\times3$ identity matrix, and $\bm{B}$ is the $3\times3$ matrix of structural paths; $\bm{\Phi}$ is a $3\times3$ matrix containing the variances and covariances of the exogenous LV as well as the residual variances and covariances of the endogenous LVs, and $\bm{\Psi}$ is a diagonal $13\times13$ matrix of the MV unique variances. The exogenous and mediating LVs each have four MVs, and the endogenous LV has five MVs. All 13 factor loadings in $\bm{\Lambda}$ were specified to be close to 0.8 to avoid equivalent population parameter values, and the matrix $\bm{\Phi}$ is specified to be diagonal with unit variances.
\\
    \textbf{Magnitude of correlations}. The magnitude of correlations among MVs was determined by either altering the size of the unique MV variances in $\bm{\Psi}$ or the size of the structural paths in $\bm{B}$. These two approaches were examined because some substantive research is solely focused on measurement models while others focus on structural paths.
\\
    \textit{\textbf{Unique variances}}. The size of unique variances of MVs indirectly affects their correlations. The variance of each MV is due to three sources of variation: common variance, specific variance, and error variance. Common variance is due to the common factor or LV, specific variance represents systematic factors affecting the MV, and error variance represents random error of measurement or unreliability. Reliable variance in an MV is the sum of common and specific variances, and unique variance is the sum of specific and error variances. Small unique variances translate to larger correlations between MVs by reflecting accuracy of measurement associated with increased power to reject false models \cite{browne.et.al.2002}. Two levels of unique variances are specified, resulting in low and high correlations among MVs. Large and small measurement errors are defined by unique variances $\bm{\Psi}_{jj}$ close to 0.5 and 0.1, respectively, where $j = 1,\ldots,13$. Small levels of random noise were added to 0.5 or 0.1 to avoid equivalent population parameter values.
\\
    \textit{\textbf{Structural paths}}. The magnitude of correlations among MVs are also influenced by the size of structural pathways in $\bm{B}$; the larger the structural paths, the larger the correlations among the MVs. Small and large structural paths were, respectively, specified as
\begin{align*}
  \bm{B}=\left(%
        \begin{array}{ccc}
        0 & 0 & 0 \\
        0.2076 & 0 & 0 \\
        -0.1989 & 0.2015 & 0 \\
        \end{array}%
        \right)
        &\mbox{    and       }
          \bm{B}=\left(%
        \begin{array}{ccc}
        0 & 0 & 0 \\
        0.6076 & 0 & 0 \\
        -0.5989 & 0.6015 & 0 \\
        \end{array}%
        \right).
\end{align*}
\\
    The 2 (size of unique variances) $\times$ 2 (size of structural paths)= 4 different models varied in the average strength of correlations among MVs. The model with large unique variances and small structural paths ($\bm{\Sigma}_1$) had the smallest average correlations followed by small unique variances and small structural paths ($\bm{\Sigma}_2$), then large unique variances and large structural paths ($\bm{\Sigma}_3$), and finally small unique variances and large structural paths ($\bm{\Sigma}_4$). Three levels of model fit ($\epsilon =$ 0, 0.03, and 0.09) where then applied to these four matrices, resulting in 12 population generating models.
\\
\vspace{12pt}
    \noindent\textbf{Data Generation and Analysis}\\
    Given the 12 population generating models, a random sample for two levels of sample size ($N=200$ and 1000) were drawn, resulting in 24 sample covariances and a total of 36 conditions including the 12 population covariances. Following convention, we constructed 95\% CRs for two focal parameters on the sample data. The FPE contours for the same two focal parameters were defined by a perturbation to the RMSEA of $\tilde{\epsilon} = .005$ (\citeNP{maccallum.browne&lee.2009}, \citeyearNP{maccallum.lee&browne.2010}) and a perturbation of $\delta_F = .05\hat{F}$, and FPEs are obtained for the population and sample covariances. Note, FPEs based on $\delta_F$ are not obtained for population covariance matrices where model fit is perfect (i.e., $F = 0$). Each CR and FPE contour was constructed with 100 points which sampled the $k_f=2$-dimensional profile likelihood surface using a variant of the root finding algorithm of \citeA{maccallum.lee&browne.2010}.
\\
    The focal parameters of interest are the two structural paths of the indirect effect $\bm{\theta}_f=(\beta_{21}, \beta_{32})'$. To identify the model with $k=29$ parameters and $df=62$, all LVs were scaled to have 1.0 variance or residual variances. This specification of the model results in profile likelihood surfaces which tend to be elliptical in nature (cf., standardized structural effects in the empirical examples to follow) such that the widths of the major and minor axes of the CRs and FPEs can be reasonably computed to numerically quantify their size. Table \ref{table:delta} presents model fit information, in terms of RMSEA and $F$, for the 36 covariances. Across the population covariances, $\epsilon$ and $F$ are invariant. In sample covariances, $\hat{\epsilon}$ and $\hat{F}$ vary between different models, sample sizes, and model fit, reflecting sampling variability and model error. As sample size decreases or model misfit increases, $\hat{F}$ increases. Similarly, $\hat{\epsilon}$ increases with increasing model misfit while holding sample size constant, and vice-versa. Note that $\hat{\epsilon}$ is slightly smaller for $N=200$ compared to $N=1000$ because of the larger effect of the sample size correction in smaller samples.
\\
\vspace{12pt}
    \noindent \textbf{Results}\\
    \textbf{Characteristics of different perturbation schemes}. Under the general perturbation framework, the first step to computing CSs or FPEs is to perturb $F$ in the population or $\hat{F}$ in the sample (see Equation \ref{eq:8}). Graphically, the perturbation $\delta$ is the vertical distance from the ML estimates of the focal parameters, $\hat{\bm{\theta}}_{0f}$, which defines CSs and FPEs; larger $\delta$ results in moving farther down the $k_f$-dimensional profile likelihood surface which leads to wider CSs and FPEs and vice-versa. In the scale of $F$, the perturbations for CSs when $N=200$ and $1000$ are fixed at .030 and .006, respectively. As expected, increasing sample sizes decreases $\delta_{CS}$ resulting in tighter CSs.
\\
    Table \ref{table:delta} presents FPE perturbations where $\tilde{\epsilon} = .005$ and $\delta_F=.05F$ in the scale of $F$, denoted by $\delta_{\epsilon}$ and $\delta_F$ respectively. From Table \ref{table:delta}, controlling for model fit, $\delta_{\epsilon}$ for the population is invariant across the four population covariances because no sample size adjustment is involved. In contrast, holding model misfit constant, $\delta_\epsilon$ varies across sample covariances and sample size because of sampling variability and the sample size adjustment. Thus, a fixed perturbation value of $\tilde{\epsilon}$ does not result in an invariant magnitude of $\delta_\epsilon$ across different populations coupled with sampling variability. Note also that when sample size is held constant, $\delta_F$ is relatively consistent across different population covariances and levels of model misfit. In general, $\tilde{\epsilon} = .005$ is associated with a larger perturbation compared to $\delta_F=.05\hat{F}$; the converse is true when sample size is small (see Equation \ref{eq:e_vs_f}).
\\
    \textbf{Sample size}. Table \ref{table:widths} presents the means and standard deviations ($SD$) of major and minor axis widths for CRs across the three levels of model fit. As expected, CRs increase in size when sample size decreases as indicated by the larger widths at smaller sample sizes; the zero $SD$s of the major and minor axes of the CRs indicate that model fit did not affect their size. Additionally, Table \ref{table:widths} also presents values of major and minor axis widths of FPE sets by sample size and model fit. For FPEs defined by RMSEA ($\tilde{\epsilon}$), increases in sample size decreased the size of FPE sets in the condition of perfect fit ($\epsilon=0$); for the conditions of imperfect fit, sample size exerted little influence on the size of sets of FPEs (see Equation \ref{eq: delta_e}). For FPEs defined by a proportion to $\hat{F}$ ($\delta_F$), increasing sample size led to smaller sets of FPEs.
    \\
    \textbf{Model fit}. Contrary to expectations, CSs were unrelated to model fit and were essentially constant in size across the three levels of model fit as evidenced by the zero $SD$s in Table \ref{table:widths}. Figure \ref{fig:sim} presents CRs and FPE contours based on $\tilde{\epsilon}$ for the 36 conditions. Figure \ref{fig:dF} presents FPEs based on $\delta_F$ for the 36 conditions. Sample size increases from the top to the middle row of plots within each figure, and ML estimates (or population parameters) for each of the four models are presented as solid geometric shapes within each plot. The last row of plots in Figures \ref{fig:sim} and \ref{fig:dF} relate to population models where only FPEs are computed. Boundary points forming CRs and FPE contours of the focal parameters are represented by analogous open geometric shapes; the large unique variances and small structural paths model ($\bm{\Sigma}_1$) is represented by squares, the small unique variances and small structural paths model ($\bm{\Sigma}_2$) is represented by triangles, the large unique variances and large structural paths model ($\bm{\Sigma}_3$) is represented by diamonds, and finally the small unique variances and large structural paths model ($\bm{\Sigma}_4$) is represented by circles. Because CSs are not influenced by model fit, they were not presented by different levels of model fit.
\\
     FPE contours were influenced by model fit as shown in Figures \ref{fig:sim} and \ref{fig:dF}. Among the plots of FPEs, model fit decreases from the left most to the right most columns of plots in each Figure. For FPEs defined by $\delta_F$ (see Figure \ref{fig:dF}), improving model fit is associated with smaller FPE contours. For FPEs defined by $\tilde{\epsilon}$, at the level of the population and at $N=1000$ (second and third rows of FPE contours in Figure \ref{fig:sim}), improving model fit is related to smaller FPE contours. However, at $N=200$, model fit does not show a monotonic relationship with the size of FPE contours (first row of FPE contours in Figure \ref{fig:sim}); the smallest FPE contour is associated with good model fit ($\epsilon=.03$), followed by perfect model fit ($\epsilon=0$), and finally poor model fit ($\epsilon=.09$). These results are not surprising. As shown in Equation~\ref{eq: delta_e}, when $\hat{\epsilon}>0$, the $\delta_{\epsilon}$ decreases with improving model fit until it reaches the minimum of $df\tilde\epsilon^2$ at $\hat{\epsilon}=0$; after that it increases with improving model fit until it reaches $df(\tilde{\epsilon}^2+(N-1)^{-1})$ at $\hat{F}=0$. Although the perturbation defining FPEs, $\delta_{\epsilon} = .005$, is a fixed value in the scale of RMSEA, this perturbation translates to different values in the scale of $F$ (i.e., $\delta_\epsilon$; see Table \ref{table:delta}), which depends on sample size and model error (see Equation 4; cf., \citeNP{chen.et.al.2008}). For instance, for the model with small unique variances and large structural paths ($\bm{\Sigma}_4$), $\tilde{\epsilon} = .005$ translates to $\delta_{\epsilon}$ = .030, .014, and .055 for perfect, good, and poor model fit in the scale of $F$ at $N=200$, respectively. For the same model, $\delta_{\epsilon}$ = .009, .018, and .056 at $N=1000$ for perfect, good, and poor model fit, respectively.
\\
    \textbf{Magnitude of correlations}. Recall that the magnitude of correlations were manipulated by either changing the size of unique variances of the MVs or the size of the structural paths, resulting in four models. Computations for these four models are presented within each plot in Figures \ref{fig:sim} and \ref{fig:dF}; the two elliptical forms lying to the bottom left of each plot represent models with small structural paths whereas the two elliptical forms lying to the top right of each plot represent models with large structural paths. The overlapping pairs of elliptical forms represent models with small and large unique variances. Note that the values of of the focal parameters in the population are identical, and increase in variability as seen by their separation in Figures \ref{fig:sim} and \ref{fig:dF} as sample size decreases.
\\
    \textbf{\textit{Unique variances}}. Holding the size of the structural paths constant, models with small unique variances are associated with smaller CSs and FPE contours for the two focal parameters compared to large unique variances ($\bm{\Sigma}_2$ versus $\bm{\Sigma}_1$ and $\bm{\Sigma}_4$ versus $\bm{\Sigma}_3$) as shown in Figures \ref{fig:sim} and \ref{fig:dF} and corroborated in Table \ref{table:widths}.  Additionally, Figures \ref{fig:sim} and \ref{fig:dF} suggest that changing the magnitude of the unique variances did not seem to change the shape of the profile likelihood surface.  As hypothesized, larger correlations due to smaller unique variances led to more peaked profile likelihood surfaces, resulting in tighter CRs and FPE contours.
\\
    \textbf{\textit{Structural paths}}. Holding unique variances constant, CRs and FPE contours for models with smaller structural paths tend to be smaller compared to those for larger structural paths in a mediation model ($\bm{\Sigma}_1$ versus $\bm{\Sigma}_3$ and $\bm{\Sigma}_2$ versus $\bm{\Sigma}_4$). From Table \ref{table:widths}, the major and minor axes of CRs and FPE contours for the two focal parameters across the different sample sizes confirm these observations. Increasing the size of structural paths increased the correlations among the MVs, but resulted in less peaked profile likelihood surfaces and larger CRs and FPE contours. Additionally, from Figures \ref{fig:sim} and \ref{fig:dF}, increasing the magnitude of structural effects also changed the shape of the likelihood surface for the focal parameters.
    \\
    These results suggest that manipulating the size of the correlations among the MVs changes the peakedness and shape of the profile likelihood surface, influencing the size and shape of CRs and FPE contours. However, it is not the magnitude of the correlations per se that determine the shape of the profile likelihood surface, but how the correlations were manipulated. Decreasing unique variances and decreasing structural effects in a mediation model led to more peaked likelihood surfaces of the focal parameters and tighter CRs and FPE contours. In contrast, increasing unique variances and increasing structural effects in a mediation model results in flatter profile likelihood surfaces and larger CRs and FPE contours.
\\
\vspace{12pt}
    \begin{centering}
    \textbf{Empirical Data Illustrations}\\
    \end{centering}
\vspace{12pt}

    \blue{Given the distinct properties of CSs and EWs, we turn to illustrating their utility and interpretation with three empirical examples below.} These examples involve using a SEM to model attitudes in prediction of initiating a conversation with a stranger (Example 1, \citeNP{bagozzi&warshaw.1988}), a regression model involving latent variables in predicting posttraumatric growth in the context of pediatric palliative care (Example 2, \citeNP{cadell.et.al.2014}), and a latent variable mediation model examining the indirect effect of adolescent spirituality on thriving through religiosity (Example 3, \citeNP{dowling.et.al.2004}). OpenMx code of examples are provided in the supplementary material.

\vspace{12pt}
    \noindent \textbf{Example 1: Conversation with a Stranger}\\

    Bagozzi and Yi \citeyear{bagozzi&yi.1988} presented research by Bagozzi and Warshaw \citeyear{bagozzi&warshaw.1988} which investigated what motivates people to initiate a conversation with an attractive stranger. Attitude towards trying to initiate such a conversation is hypothesized as a predictor of actual, subsequent Trying to initiate a conversation in the succeeding week, and Intention to try is included as a mediator. Figure \ref{fig:ex1} presents a path diagram of the hypothesized relationship which also includes Subjective Norm of others' opinions that one should try as a covariate. Attitude and Intention are measured by three and two indicators, respectively. Trying and Subjective Norm are measured variables themselves. The model has a total of 17 parameters, and the correlation matrix of all measured variables based on $N=250$ is reported in Table 5 of Bagozzi and Yi \citeyear{bagozzi&yi.1988}.

    Our analysis reproduced the estimates of model fit reported in Bagozzi and Yi \citeyear{bagozzi&yi.1988} with $\chi^2(11) = 6.801$, $\hat{F}=0.0273$, RMSEA = 0, and CFI = TLI = 1. Parameter estimates of all parameters are given in the supplementary material. Below, we focus on the direct effect of Attitude on Trying and its indirect effect through Intention. Parameter estimates, CIs and FPEs of these parameters are summarized in Table~\ref{tab:stranger example}. We first consider the direct effect. The ML estimate of this path is very small with a $95\%$ profile likelihood CI including zero, suggesting a lack of evidence of the existence of a direct effect beyond the mediation effect through Intention \blue{in the population}. The perturbation applied to $\hat{F}$ is $\delta_{CS} = \chi^2_{.95, 1}/(250-1) = 0.0154$ for this CI.

    From a sensitivity analysis, two perturbations to model fit were first selected for RMSEA, $\tilde\epsilon_1 = .001$ ($\delta_{FPE_1} = .0169$) and $\tilde\epsilon_2 = .005$ ($\delta_{FPE_2} = .0171$). The two pairs of FPEs coincide with the CI to the second decimal place because the three perturbations are very close in magnitude. The FPEs suggest that slight changes in model fit could change the sign of the direct effect, \blue{raising the question of whether inferences based on the parameter estimate are valid}. We further perturb the model fit with $\delta_{FPE_3} = 0.02\hat{F} = 0.00055$ and $\delta_{FPE_4} = 0.05\hat{F} = 0.00137$. Neither pairs of FPEs includes zero, indicating that the sign of the estimate is \blue{robust} to these perturbations. The lack of evidence of a direct effect is entirely due to sampling error.

\begin{figure}[!htbp]
\centering
\includegraphics[width=6in]{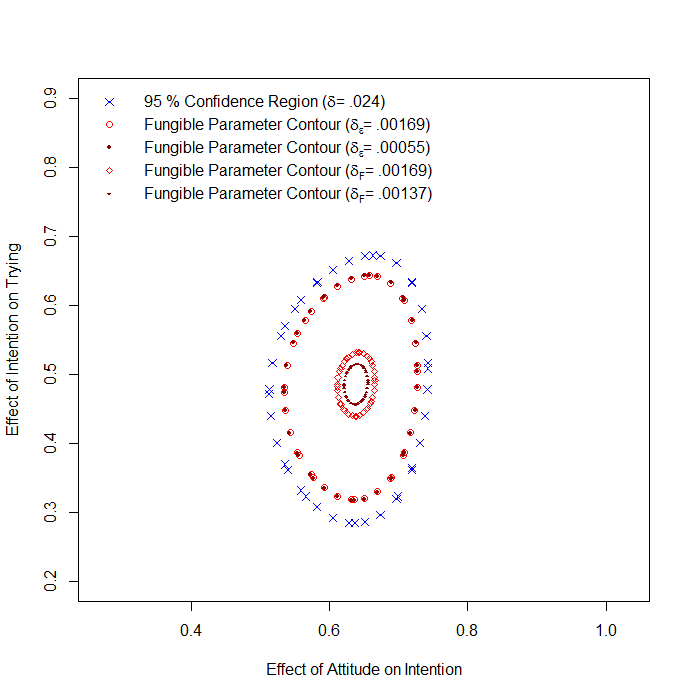}
 \caption{Profile likelihood confidence region and fungible parameter contours for the indirect effect of Attitude on Trying. }
 \label{fig:bagozzi}
\centering
\end{figure}

    We now consider the indirect effect of Attitude on Trying through Intention. The ML estimates for both paths are positive. A perturbation of $\delta_{CS} = \chi^2_{.95,2}/(250-1) = 0.024$ defines the $95\%$ joint CR. The same four perturbations defined above were used to compute four sets of FPEs. Because the perturbation defining the CR is now larger than those defining the FPEs, the CR is larger than the FPE contours. The intervals listed in Table~\ref{tab:stranger example} are projections of the CR and sets of FPEs from two dimensions to one dimension. Figure \ref{fig:bagozzi} depicts the CR and FPE contours. The largest contour of crosses represent the CR. The next two largest contours, represented by open and closed circles, are the two sets of overlapping FPEs obtained from perturbing the RMSEA. Finally, the two smallest FPE contours, represented by open and closed diamonds, are associated with perturbations of $\hat{F}$. The 95\% CR does not contain any values of 0, implying that the two estimated positive effects are both significant at the 5\% level of significance. The four sets of FPE contours are even smaller, suggesting the estimates are stable against slight perturbations to RMSEA and $\hat{F}$, \blue{justifying the valid interpretation and inference afforded by these parameter estimates}.

    This example illustrates the situation where FPEs are smaller than or of about equal size to a CS. In this situation, parameter estimates are not as sensitive to perturbations to model fit as they are to sampling error, and conclusions based entirely on the analysis of sampling variability is valid.
 \\
 \vspace{12pt}
    \noindent \textbf{Example 2: Posttraumatic Growth of Caregivers}\\
   \citeA{cadell.et.al.2014} investigated factors which contribute to post-traumatic growth (PTG) of caregivers to children with a life-limiting illness. In this context of caregiving, PTG refers to the positive changes that people experience as a result of adverse circumstances due to a traumatic event \cite{tedeschi.pak&calhoun.1998} and is measured by five subscales of the Posttraumatic Growth Inventory (PTGI; \citeNP{tedeschi&calhoun.1996}): Relating to Others ($y_1$), New Possibilities ($y_2$), Personal Strength ($y_3$), Appreciation of Life ($y_4$) and Spiritual Change ($y_5$). Because ``finding meaning'' is an important part of the process which leads to PTG, meaning in caregiving (MCG) that is defined as the sense which people make of their caregiving experiences is included as a key variable. Five additional variables were also included in their study: Self-esteem, Optimism, Spirituality, Depression and Caregiver burden. The correlation matrix and standard deviations of the five subscales of PTGI and the six other variables are presented in Table 2 of \citeA{cadell.et.al.2014}, and the total sample size is $N = 273$.
 	
   For illustrative purposes, we consider a latent variable regression model where PTG is predicted by MCG, Spirituality, and a latent variable Psychological Well-being (PWB) which is indicated by Self-esteem ($x_1$), Optimism ($x_2$), Depression ($x_3$) and Caregiver Burden ($x_4$). MCG and Spirituality are single indicators and the square root of their reliability values ($0.82$ and $0.93$ as reported in \citeNP{cadell.et.al.2014}) were used as fixed (standardized) loadings. All variables are standardized, with 11 manifest variable standard deviations included separately as nuisance parameters. The model has a total 27 parameters, including a free path from spirituality to the Spiritual Change subscale of PTG (see Figure \ref{fig:ex2}). This model has acceptable fit, with $\chi^2(39)=112.71$, $\hat{F}=0.414$, RMSEA$= 0.0834$ with $95\%$ CI $(0.062,0.105)$, CFI$ = 0.948$ and TLI$ = 0.927$. All loadings are above $0.40$ and significant at $p < .05$. Details of parameter estimates are reported in the supplementary material. We focus on the regression paths as presented in Table \ref{tab:PTG example}.
 	 	
   	Parameter estimates, CIs and FPEs are summarized in Table~\ref{tab:PTG example}. The point estimates show a large effect of MCG on PTG, controlling for Spirituality and PWB. The regression coefficients associated with Spirituality and PWB are negative, implying that an increase in one of these variables predicts a decrease in PTG after controlling for MCG and the other variable. The $95\%$ point-wise CIs suggest that the effect of MCG on PTG is large and statistically significant, whereas the conditional effects of Spirituality and PWB on PTG could be non-existent in the population because their CIs are close to zero. These CIs correspond to the perturbation to $\hat{F}$, $\delta_{CS} = \chi^2_{.95, 1}/(273-1) = 0.0141$.
 	
   	 Next, we compute FPEs for the three focal coefficients in a sensitivity analysis. The same four perturbations applied in Example 1 were also applied to Example 2, and the resultant increases in $\hat{F}$ are presented in the last column of Table~\ref{tab:PTG example}. The first and third pairs of FPEs (i.e., based on $\tilde{\epsilon}_1$ and $\delta_{F_1}$) are tighter around their ML estimates than the CIs because $\delta_{FPE} <\delta_{CS}$. For these $\delta_{FPE}$ perturbations, parameter estimate uncertainty is smaller than that due to sampling error. In contrast, the second and fourth FPEs (i.e., based on $\tilde{\epsilon}_2$ and $\delta_{F_2}$) are wider than the CIs. Note, the most variable pair of FPEs (defined by $\tilde{\epsilon}_2$) is associated with negative and positive alternative values for both individual effects of Spirituality and PWB on PTG, suggesting that these individual effects are not robust. Taken together, the sensitivity analysis regarding the stability of individual coefficients suggests that the individual conditional effects of Spirituality and PWB on PTG are both unstable, \blue{raising questions on whether inferences about these effects are valid}.
 	
   In addition to individual coefficients, we can also consider the total effect of Spirituality and PWB, controlling for MCG, on PTG. This contribution can be measured as the variance of PTG due to Spirituality and PWB for a given level of MCG, which is a function of the path coefficients and the correlations among the three predictors. The point estimate of this variance, $0.029$, is relatively small; although relatively close to zero, the $95\%$ CI $(0.004,0.075)$ suggests that this effect is significant at the 5\% level of significance. Sensitivity analysis provides information on the lack of \blue{validity} of this result in that this effect could essentially be zero (i.e., $3\times 10^{-6}$) based on largest of the four perturbations. \blue{The FPEs suggest that inferences about this total effect are potentially invalid.}
 	
 	This example demonstrates the unique utility of FPEs as a \blue{diagnostic device for evaluating whether parameter estimates can be validly interpreted based on their stability in relation to the model's fit to data}. For a model whose fit is acceptable but not close, perturbations to model fit may reveal instability in parameter estimates which exceed the uncertainty due to sampling error even for medium sample size. Thus, inspection of FPEs would provide valuable diagnostic information unavailable from CIs and CRs.
 	
\vspace{12pt}
 	\noindent \textbf{Example 3: Adolescent Thriving}\\

  \citeA{dowling.et.al.2004} investigated the relationship among Spirituality, Religiosity, and Thriving of adolescents in a latent variable mediation model. Thriving is defined as positively developing, Religiosity is defined as the relationship between a particular doctrine about a supernatural power through institutional affiliation and participation in prescribed practices \cite{reich.oser&scarlett.1999}, and Spirituality is defined as seeing life and living in new and better ways and taking something to be transcendant \cite{reinhart.2015}. It was hypothesized that the effect of Spirituality on Thriving is mediated through Religiosity, and data from a sample of $N=1000$ youth from the Search Institute Young Adolescents and their Parents archival data set was analyzed. Religiosity, Spirituality, and Thriving were conceptualized as second order factors (e.g., see \citeNP{dowling.et.al.2003}), each subsuming four, three and nine first order factors, respectively; each first order factor was indicated by two or three items, with a total of $47$ items. Despite a second order factor conceptualization, the authors reported a model with the three latent variables measured directly by the items without the first order factors. They reported that their model fit well, $\chi^2(998)=2441.28$, RMSEA$ = 0.049$ and CFI $= 0.90$, and the indirect effect of Spirituality on Thriving through Religiosity was significant. 

  For illustrative purposes, we simplified the model and selected 14 out of the 47 variables to indicate the three latent variables. One variable was selected from each of the 16 lower order factors to maximize content validity of the latent variable save for ``Participation in Activities of Self-Interest'' because items loaded weakly ($<0.1$) onto this factor. Additionally, only one of the six items measuring the lower order factors of ``Rules for Youth Presented by Father" and ``Rules for Youth Presented by Mother'', which are subsumed by Thriving, was selected to avoid correlated residuals. This model has 31 parameters and yields $\chi^2(74)= 254.7$, RMSEA $= 0.049$, CFI $=0.87$, TLI $= 0.84$ and SRMR $=0.042$. Parameter estimates of all parameters are reported in the accompanying supplementary material. Below we focus on the latent variable mediation.
	
  Parameter estimates, CIs and FPEs of the three paths among the latent variables are summarized in Table~\ref{tab:thriving example}. The 95\% point-wise CI for the direct effect is constructed with $\delta_{CI} = \chi^2_{.95, 1}/(1000-1) = 0.0038$; the 95\% joint CIs for the two paths of the indirect effect are constructed with $\delta_{CI} = \chi^2_{.95, 2}/(1000-1) = 0.0060$. The first, third and fourth FPEs were constructed with the same perturbations as in the other two examples. Because $\tilde\epsilon_2 = .005$ ($\delta_{FPE_2} = .0384$) would lead to very wide FPEs which require a delicate choice of parameter boundaries in the code for proper convergence, the second FPE was defined as $\delta_{FPE_2} = .0300$ ($\tilde\epsilon_2 = .00394$) instead.
	
\begin{figure}[!htbp]
\centering
\includegraphics[trim=0cm 0cm 0cm 1.5cm, clip=true,width=4.6in]{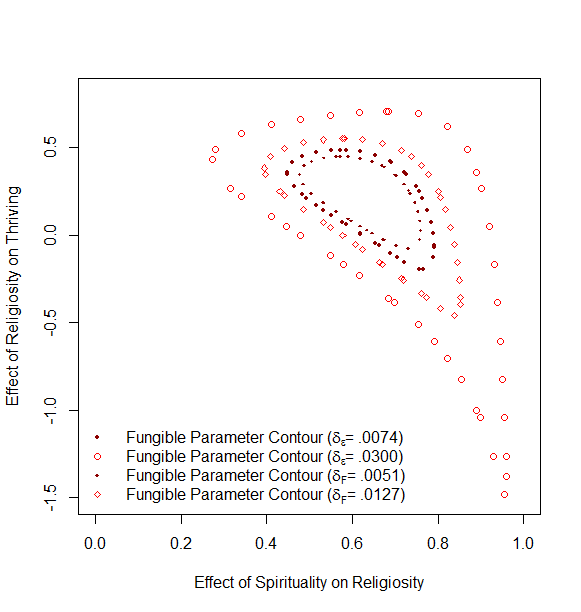}
\includegraphics[trim=0cm 0cm 0cm 1.5cm, clip=true,width=4.6in]{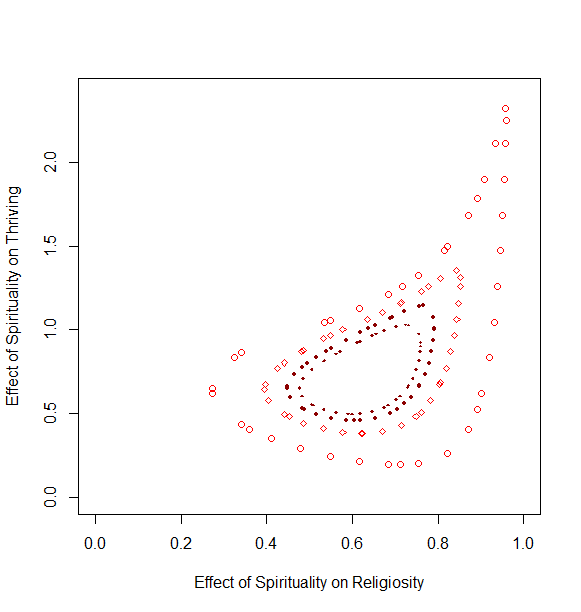}
 \caption{Fungible parameter contours for Example 3. }
 \label{fig:dowling}
\end{figure}

  All parameter estimates of the three paths are positive, but the 95\% CI of one of the indirect path includes zero, suggesting that the positive indirect effect of Spirituality on Thriving through Religiosity could be due to sampling error. The CIs for the direct effect is smaller than the four pairs of FPEs, and the CIs for the two paths of the indirect effect are smaller than three of the four FPEs. The small CIs are due to the large sample size. Although the two smaller pairs of FPEs are similar to the CIs, the two larger pairs of FPEs show greater instability in the values with which the focal parameters can take. In particular, the second set of FPEs defined by $\tilde{\epsilon}_2$ implies a direct path greater than $2$ and an indirect path of almost $-1.5$. Figure~\ref{fig:dowling} presents plots of FPE contours for two pairs of the three paths. They convey a similar message in that the largest FPE contour  (represented by open circles) communicates instability of the focal effects predicting Thriving from Religiosity and Spirituality, \blue{raising concerns regarding their validity}.

  A closer look at this FPE contour reveals that this is due to collinearity: an alternative value of the effect of Spirituality on Religiosity is $0.96$, which is extremely close to 1.0. In this simple linear regression of Religiosity on Spirituality, because the standardized regression coefficient is also the correlation coefficient, these two variables as predictors of Thriving could become highly correlated and manifest as a problem of collinearity. Indeed,  Figure \ref{fig:dowling} shows that the extreme FPE values of the coefficients of Spirituality and Religiosity (on the ordinate) are associated with this near perfect correlation between the two predictors (on the abscissa). Unlike wide CIs which can be minimized by a larger sample size, the variability in the sets of FPEs suggest a possible issue with the research design. For a given correlation between Spirituality and Religiosity in the population, its FPEs could be reduced by the use of indicators with higher loadings (i.e., improved measurement) and a model which fits better (see targeted illustration above). \blue{This problem of collinearity which invalidates the results would have been inappropriately overlooked without the consideration of FPEs.}
	
  This example again shows the unique value of FPEs in a sensitivity analysis. When sample size is large, CIs and CRs are usually small due to the limited sampling error. In such situations, small perturbations to model fit may reveal instability in parameter estimates which exceed the uncertainty due to sampling error. Thus, inspection of FPEs could provide valuable diagnostic information \blue{regarding the valid interpretation of parameter estimates which are} unavailable from CIs and CRs.

\vspace{12pt}
    \begin{centering}
    \textbf{Summary and Discussion}\\
    \end{centering}
\vspace{12pt}
    Confidence sets and FPEs communicate different aspects of parameter uncertainty which provide information with regards to the extent to which a scientific finding can be rigorously interpreted. Confidence sets communicate sampling variability and estimate precision whereas FPEs communicate \blue{information on whether parameter estimates can be validly interpreted based on} the stability of parameter values in relation to \blue{the specified model's fit to data}. By introducing a perturbation framework based on the likelihood function, it is shown that CSs and FPEs share similar properties despite their distinct interpretations.  Given their apparent commonalities, we clarify the theoretical relationship between CSs and FPEs by establishing their analytical relationship. Indeed, CSs and FPEs are horizontal intersections of the likelihood surface defined by the vertical distance or perturbation $\delta$ from the maximum of the likelihood.  We demonstrate distinct characteristics between CSs and FPEs in a targeted illustration and summarize key differences between CSs and FPEs in Table \ref{table:sum}. Further, we illustrate with three empirical examples the value of considering the distinct information communicated by CSs and FPEs regarding the defensibility of estimated parameter estimates or effect sizes (cf.,\citeNP{green.1977}).
\\
\vspace{12pt}
    \noindent\textbf{Theoretical Properties}\\
    In general, the size of CSs and FPE sets are influenced by: (a) the magnitude of the perturbation $\delta$ that uniquely defines these two kinds of parameter uncertainty, or (b) the peakedness and shape of the likelihood surface. Note that when sample covariances are modeled, the likelihood surface is influenced by sampling variability.
\\
    \textbf{Perturbations}. Profile likelihood CSs are defined by a perturbation $\delta_{CS}$ that is determined by sample size, the Type I error rate $\alpha$, and the $\chi^2$ quantile. Larger sample sizes, larger error rates, and smaller quantiles that are determined by a smaller number of $k_f$ focal parameters, result in smaller perturbations and tighter CSs. Conversely, smaller sample sizes, smaller $\alpha$  levels, and larger quantiles define larger CSs. The perturbation defining CSs is primarily a function of sample size and parameter degrees of freedom, and independent of model fit; thus, CSs carry information only about sampling variability and not model fit.
\\
    Fungible parameter estimates are defined by a perturbation, chosen by the researcher, such that the perturbed model fit is practically no different from optimal model fit. Larger perturbations, in the scale of $F$, result in larger FPE contours and vice-versa. Based on current convention (\citeNP{lee.maccallum&browne.inpress}; \citeNP{maccallum.lee&browne.2013}; \citeNP{maccallum.browne&lee.2009}), we examined perturbations that define FPEs in the scale of the RMSEA, $\tilde{\epsilon}$ and introduce an alternative perturbation as a percentage of $\hat{F}$. For a given $\tilde{\epsilon}$, the size of this perturbation in the scale of $F$ is primarily determined by model fit, and is not very much affected by sample size for a fixed level of model fit in the population when $\hat{\epsilon}>0$. Only when $\epsilon=0$ in the population does increasing sample size show a decreasing effect on the size of sets of FPEs defined by $\tilde{\epsilon}$. Alternatively, FPEs defined by $\delta_F$ increase in size with decreasing model fit. Unlike FPEs defined by $\tilde{\epsilon}$, sample size has a negative effect on the size of FPE sets defined by $\delta_F$ for a fixed level of model fit in the population but has no effect for a fixed level of model fit in a sample. However, $\delta_\epsilon$ and $\delta_F$ are monotonically related in certain conditions (see Equation \ref{eq:e_vs_f}). In general, FPE contours carry information largely about model fit, which dominates information about sampling variability.
\\
    \textbf{Likelihood surface}. The likelihood function fully represents parameter uncertainty \cite{pawitan.2001}, and CSs and FPEs are horizontal `slices' of the likelihood surface which are determined by their vertical distances ($\delta_{CS}$ and $\delta_{FPE}$) from the maximum point of the likelihood surface, $\hat{\bm{\theta}}$. The peakedness and shape of the likelihood surface thus determines the size and form of CSs and FPE contours or surfaces. Contrary to expectations, changes in model fit via the Cudeck and Browne \citeyear{cudeck&browne.1992} method do not alter the shape of the likelihood surface, and changes in the magnitude of the correlations among the MVs do not directly change the shape of the likelihood surface. Instead, manipulating the unique variances of MVs results in different levels of likelihood surface peakedness; smaller unique variances reflecting higher measurement reliability have more peaked likelihood surfaces, whereas larger unique variances reflecting higher measurement error have less peaked likelihood surfaces. Additionally, changing the magnitude of structural paths (in a latent variable mediation model) results in changing the shape and peakedness of the likelihood surface; larger structural paths were associated with shallower likelihood surfaces that tended to detract from an elliptical form whereas smaller structural paths had more peaked likelihood surfaces that were more elliptical in shape.
\\
\vspace{12pt}
    \noindent\textbf{Considerations for Practice}\\
    Confidence sets and FPEs are distinct kinds of parameter uncertainty which provide unique and useful information to substantive researchers in terms of \blue{validly and} rigorously interpreting their effect sizes. Table \ref{table:sum} provides a summary of their distinct properties and uses. In any data analysis, diagnostics should be conducted to ensure the \blue{validity} and stability of the model solution (cf., \citeNP{belsey.kuh&welsch1980}; \citeNP{cook&weisberg.1982}; e.g., \citeNP{pek&maccallum.2011}). Scientific conclusions based on inferential methods should move away from null hypothesis significance tests which tend to result in essentialist dichotomies (i.e., significant or not; see \citeNP{cohen.1994}). By considering parameter uncertainty due to sampling variability as quantified by CSs, and slight perturbations to model fit as quantified by sets of FPEs, results of an analysis can more seamlessly be interpreted to reflect degrees of confidence in their defensibility.
\\
    A model which can appropriately describe or explain phenomena well should fit the data well, be robust in its description of the data as reflected by the size of FPE contours or surfaces \blue{such that parameter estimates can be validly interpreted, and have tight CSs} \cite{green.1977}.\footnote{The link between sample size and parameter sensitivity has been made by \citeA{davis-stober.2011} for multiple linear regression.} Confidence sets provide a measure of uncertainty in the estimation of parameters due to sampling variability; they inform researchers of the statistical significance of parameter estimates, the precision and stability of estimation, and provide a range of plausible population parameter values. Tight CSs provide a strong basis for inferences whereas wide CSs weaken confidence in making inferences about the population. The information inherent in CSs are strictly about statistical inference. On the other hand, FPEs convey \blue{diagnostic information regarding the validity} of the model's description of the data, afforded by optimal parameter estimates, in relation to the \blue{model's fit to data}. When slight perturbations result in compact and tight FPE contours, the model's description of the data is stable, providing a basis for \blue{interpreting optimal parameter estimates and} drawing strong and definitive conclusions. Alternatively, when large FPE contours are observed such that wildly varying alternative parameter estimates describe the data just as well as the optimal estimates in terms of model fit, this instability of the model's description of the data casts doubts on the validity behind interpreting the optimal parameter estimates. Thus, the information communicated by FPE contours can either enhance or undermine the \blue{validity} and definitive interpretation of optimal parameter estimates.
\\
    Consistent with calls to focus on the stochastic nature of statistical results (e.g., \citeNP{cumming&fidler.2009}; \citeNP{waldman&lilienfeld.2015}), we encourage researchers to report CSs about their effects or focal parameter estimates (for example, see \citeNP{steinberg&thissen.2006}). \blue{Prior to interpreting CSs, researchers should routinely conduct sensitivity analysis to assess the validity of statistical results, and FPE contours communicate the stability of a model's description of the data via parameter estimates in relation to model fit.} To that end, we provide example OpenMx code for two of our empirical examples to compute CSs and FPEs in the supplemental material. The practice of constructing CSs and FPEs convey the stochastic nature of statistical results, and can buttress the case for drawing strong scientific conclusions. Because research regarding the type and magnitude of perturbation to define FPEs is still under development, we recommend using several different magnitudes of $\tilde{\epsilon}$ and $\delta_F$ when constructing FPEs to fully explore the stability of the model's description afforded by parameter estimates under slight perturbations to model fit.
\\
\vspace{12pt}
    \noindent\textbf{Future Directions}\\
    In this paper, we consolidated contemporary work on profile likelihood CSs \cite{pek&wu.2015} and FPEs (\citeNP{lee.maccallum&browne.inpress}; \citeNP{maccallum.browne&lee.2009, maccallum.lee&browne.2010}) in SEM by establishing their analytical relationship. We introduced an alternative definition of FPEs so as to better understand the nature of FPEs in relation to the measure of model misfit to be perturbed. We also illustrated how these two kinds of parameter uncertainty are distinct, as well as their relevance to drawing definitive scientific conclusions about parameter estimates in practice. To have a fuller understanding of the nature of CSs and FPEs, other factors which impact their size and shape such as the number and types of parameters in the model should be examined in future studies. A better understanding of the factors which affect CSs and FPE contours is valuable for theory and practice because the size and shape of FPE contours and CSs determine the limits to which definitive interpretations of parameter estimates can be made. Another extension to this work involves developing a single quantification of parameter uncertainty that jointly takes into account stability of the model's description of the data via parameter estimates and sampling variability (cf., \citeNP{wu&browne.2015a,wu&browne.2015b}). Further developing methods to quantify different aspects of parameter uncertainty can only serve to facilitate researchers' informed choices behind drawing \blue{valid and} definitive scientific conclusions.
    \\
%
%
%
\vspace{12pt}
\begingroup
\renewcommand{\chapter}[1]{\noindent \textbf}%
\bibliography{ref}
\endgroup
\newpage
\begin{doublespacing}
 \begin{landscape}

 \ctable[caption = Distinctions between Confidence Sets and Fungible Parameter Estimates,
 label = table:sum,
 width = 200mm,
 pos = htb!,
 center
] {>{\raggedright}p{4cm}>{\raggedright}p{7cm}p{8cm}}{ \tnote[]{\textit{Note}. $N$=sample size and $\alpha$ = Type I error rate.}}
    {\FL
                 &Confidence Sets & Fungible Parameter Estimates \NN
                 \cmidrule(lr){2-2}\cmidrule(lr){3-3}
     Purpose     &Inference       & Diagnostics; sensitivity Analysis\\ \cmidrule(rl){1-3}
     Assumption &Data follows some distribution &None\\ \cmidrule(rl){1-3}
     Type of uncertainty & Sampling variability &Stability of parameter estimates in relation to model fit\\ \cmidrule(rl){1-3}
     Magnitude of perturbation &
     Single value determined by $N$, $\alpha$, and critical value of test to be inverted &Several arbitrary values; suboptimal fit practically the same as optimal fit\\ \cmidrule(rl){1-3}
     Region within boundary points &Associated with same level of confidence & Associated with smaller perturbation to model fit\\ \cmidrule(rl){1-3}
     Interpretation of points &Range of plausible population values for the set of focal parameter(s) &Alternative parameter estimates which explain the data just as well as the optimal parameter estimates in terms of model fit\\ \cmidrule(rl){1-3}
     Properties &Converge to a point with increasing $N$ &Converge to a set with increasing $N$\\
                &Unrelated to model fit &Smaller with improved model fit
 \LL}

   \end{landscape} 

\ctable[caption = Model Fit and Perturbation Values for FPEs in scale of $F$,
 label = table:delta,
 width = 160mm,
 pos = htb!,
 center
] {ccccccccccccc}{ \tnote[]{\textit{Note}. FPE = fungible parameter estimates, $N=$ sample size, $\epsilon$ = population model fit in the scale of RMSEA, $\hat{\epsilon}$= sample RMSEA, and $\delta_{\epsilon}$ = perturbation of $\tilde{\epsilon} = .005$ in the scale of $F$; $\%F$ = perturbation of $\delta_F = .05F$; $\bm{\Sigma}_1 = $ large unique variances and small structural effects, $\bm{\Sigma}_2 = $ small unique variances and small structural effects, $\bm{\Sigma}_3$= large unique variances and large structural effects,
and $\bm{\Sigma}_4 =$ small unique variances and large structural effects.}}
    {\FL
 &\multicolumn{4}{c}{Population} &\multicolumn{4}{c}{$N=1000$} &\multicolumn{4}{c}{$N=200$}\\
  \cmidrule(rl){2-5} \cmidrule(rl){6-9} \cmidrule(rl){10-13}
  Model &$\epsilon$ &$F$ &$\delta_{\epsilon}$ &$\delta_F$& $\hat{\epsilon}$ &$\hat{F}$ &$\delta_{\epsilon}$ &$\delta_F$ &$\hat{\epsilon}$ &$\hat{F}$ &$\delta_{\epsilon}$ &$\delta_F$\ML
 $\bm{\Sigma}_1$ &0	&0	&.002	&$-$ &0	&.059 &.004 &.003 &0	&.277 &.037 &.014\\
                &.03 &.056 &.020 &.003 &.027 &.106 &.018 &.005 &.021 &.338 &.014 &.017\\
                &.09 &.502 &.057 &.025 &.088 &.530 &.055 &.026 &.087 &.793	&.056	&.040
\ML
$\bm{\Sigma}_2$ &0	&0	&.002 &$-$ &0 &.056 &.007 &.003 &0 &.280 &.033 &.014\\
                &.03 &.056 &.020 &.003 &.028 &.110 &.019	&.005 &.021 &.338 &.014 &.017\\
                &.09 &.502 &.057 &.025 &.088 &.544 &.056 &.027 &.087 &.784 &.056 &.039\ML
$\bm{\Sigma}_3$ &0 &0 &.002 &$-$ &0 &.056 &.007 &.003 &0 &.289 &.024 &.014\\
                &.03 &.056 &.020 &.003	&.025 &.102 &.017	&.005 &.022	&.343 &.015 &.017\\
                &.09 &.502 &.057 &.025 &.086 &.526 &.055 &.026 &.087 &.781 &.055&.039 \ML
$\bm{\Sigma}_4$ &0 &0 &.002 &$-$ &0 &.054	&.009 &.003 &0 &.283 &.030 &.014\\
                &.03 &.056 &.020 &.003 &.026 &.104 &.018 &.005 &.020 &.337 &.014 &.017\\
                &.09 &.502 &.057 &.025 &.087 &.534 &.056 &.027 &.086 &.774 &.055 &.039\LL}

\begin{landscape}
\thispagestyle{plain}

\ctable[caption = Widths of Major and Minor Axes across Model Fit,
 label = table:widths,
 width = 250mm,
 pos = hbp!,
 center
] {cccccccccccccccccc}{ \tnote[]{\textit{Note}. FPE = fungible parameter estimate, $N=$ sample size, and $SD$ = standard deviation; $\bm{\Sigma}_1 = $ large unique variances and small structural effects, $\bm{\Sigma}_2 = $ small unique variances and small structural effects, $\bm{\Sigma}_3$= large unique variances and large structural effects,
and $\bm{\Sigma}_4 =$ small unique variances and large structural effects; $\epsilon$ = population value of model fit in the scale of RMSEA.}}
    {\FL
 &&\multicolumn{4}{c}{Confidence Sets} &\multicolumn{6}{c}{FPEs ($\tilde{\epsilon}=.005$)} & \multicolumn{6}{c}{FPEs ($\delta_F=.05$)}\\
  \cmidrule(rl){3-6} \cmidrule(rl){7-12} \cmidrule(rl){13-18}
 &&\multicolumn{2}{c}{Major Axis} &\multicolumn{2}{c}{Minor Axis} &\multicolumn{2}{c}{$\epsilon=0$} &\multicolumn{2}{c}{$\epsilon=.03$} &\multicolumn{2}{c}{$\epsilon=.09$} &\multicolumn{2}{c}{$\epsilon=0$} &\multicolumn{2}{c}{$\epsilon=.03$} &\multicolumn{2}{c}{$\epsilon=.09$}\\
  \cmidrule(rl){3-4} \cmidrule(rl){5-6}  \cmidrule(rl){7-8} \cmidrule(rl){9-10}  \cmidrule(rl){11-12} \cmidrule(rl){13-14} \cmidrule(rl){15-16} \cmidrule(rl){17-18}
 &$N$ &Mean &$SD$ &Mean &$SD$ &Major &Minor &Major &Minor &Major &Minor &Major &Minor &Major &Minor &Major &Minor\ML
$\bm{\Sigma}_1$ &1000 &0.19 &0  &0.18 &0 &0.16	&0.15	&0.33	&0.32	&0.59	&0.56	&0.13	&0.13	&0.18	&0.17	&0.40	&0.38 \\
                &200  &0.43 &0  &0.40 &0 &0.48	&0.44	&0.30	&0.28	&0.60	&0.55	&0.29	&0.27	&0.32	&0.30	&0.50	&0.46 \ML
$\bm{\Sigma}_2$ &1000 &0.17 &0  &0.16 &0 &0.18	&0.18	&0.29	&0.29	&0.51	&0.50	&0.11	&0.11	&0.16	&0.15	&0.36	&0.35 \\
                &200  &0.38 &0  &0.36 &0 &0.39	&0.38	&0.26	&0.25	&0.52	&0.50	&0.26	&0.25	&0.28	&0.27	&0.43	&0.42 \ML
$\bm{\Sigma}_3$ &1000 &0.25 &0  &0.20 &0 &0.27	&0.22	&0.42	&0.34	&0.75	&0.61	&0.17	&0.14	&0.23	&0.18	&0.52	&0.42 \\
                &200  &0.56 &0  &0.44 &0 &0.50	&0.39	&0.40	&0.31	&0.76	&0.59	&0.38	&0.30	&0.42	&0.33	&0.63	&0.50 \ML
$\bm{\Sigma}_4$ &1000 &0.20 &0  &0.17 &0 &0.25	&0.22	&0.35	&0.30	&0.62	&0.53	&0.14	&0.12	&0.19	&0.16	&0.43	&0.37 \\
                &200  &0.46 &0  &0.39 &0 &0.46	&0.39	&0.32	&0.27	&0.63	&0.53	&0.32	&0.27	&0.35	&0.29	&0.53	&0.45 \LL
 }
  \end{landscape}
\clearpage 

\ctable[caption = Parameter estimates{,} CIs and FPEs for Example 1,
 label = tab:stranger example,
 width = 175mm,
 pos = htb!,
 center
] {lcccc}{ \tnote[]{\textit{Note}. ML = maximum likelihood, $CI$ = confidence interval, $FPE$ = fungible parameter estimate, $df$ = degrees of freedom, and $\delta$ = perturbation in the scale of $F$. Additionally, $\tilde\epsilon$ = perturbation to the RMSEA and $\delta_F$ = perturbation to $\hat{F}$ by a percentage.}}
    {\FL
     		& Attitude $\rightarrow$ Trying & Attitude $\rightarrow$ Intention	& Intention $\rightarrow$ Trying		& $\delta$\ML
     ML Estimate	& $0.050$			& $0.639$ 		& $0.486$			&\\
     $CI$ ($df=1$)& $[-0.119,0.216]$	&				&					&$0.0154$\\
     $CI$ ($df=2$)&					&$[0.513,0.743]$&$[0.285,0.672]$	&$0.024$ 	\\
     $FPE_1$ ($\tilde\epsilon_1 = .001$)	& $(-0.126,0.223)$	&$(0.535,0.727)$&$(0.318,0.643)$	& $0.0169$\\
     $FPE_2$ ($\tilde\epsilon_2 = .005$)	& $(-0.128,0.225)$	&$(0.534,0.728)$&$(0.317,0.644)$	& $0.0171$	\\
     $FPE_3$ ($\delta_{F_1} = 0.02\hat{F}$)	& $(\ 0.019, 0.082)$	&$(0.624,0.656)$&$(0.457,0.515)$	& $0.00055$\\
     $FPE_4$ ($\delta_{F_2} = 0.05\hat{F}$)	& $(0.00027,0.100)$ &$(0.611,0.666)$&$(0.440,0.532)$	& $0.00137$ \LL}

\clearpage 
\ctable[caption = Parameter estimates{,} CIs and FPEs for Example 2,
 label = tab:PTG example,
 width = 175mm,
 pos = htb!,
 center
] {lcccc}{\tnote[]{\textit{Note}. PTG = posttraumatic growth, MCG = meaning in caregiving, and PWB = psychological well-being; ML = maximum likelihood, $CI$ = confidence interval, $FPE$ = fungible parameter estimate, $df$ = degrees of freedom, and $\delta$ = perturbation in the scale of $F$. Additionally, $\tilde\epsilon$ = perturbation to the RMSEA and $\delta_F$ = perturbation to $\hat{F}$ by a percentage.}}
  {\FL
           & MCG $\rightarrow$ PTG  &Spirituality $\rightarrow$ PTG	&PWB $\rightarrow$ PTG				& $\delta$\ML
 ML Estimate  &$0.870$ & $-0.083$ &$-0.154$ 		&\\
 $CI$ ($df=1$)  & $[0.747,0.994]$   &$[-0.219,0.049]$ &$[-0.267,-0.041]$	&$0.0141$\\
 $FPE_1$ ($\tilde\epsilon_1 = .001$)		& $(0.787,0.953)$	&$(-0.175,0.007)$  &$(-0.231,-0.077)$	& $0.0065$\\
 $FPE_2$ ($\tilde\epsilon_2 = .005$)		& $(0.677,1.067)$	&$(-0.296,\ 0.120)$  &$(-0.330,0.020)$	& $0.0335$\\
 $FPE_3$ ($\delta_{F_1} = 0.02\hat{F}$)	& $(0.777,0.964)$	&$(-0.186,0.018)$  &$(-0.240,-0.068)$	& $0.0083$\\
 $FPE_4$ ($\delta_{F_2} = 0.05\hat{F}$)		& $(0.720,1.021)$   &$(-0.248,0.077)$  &$(-0.292,-0.018)$	& $0.0207$\LL}

 \clearpage 
\begin{landscape}
\ctable[caption = Parameter estimates{,} CIs and FPEs for Example 3,
 label = tab:thriving example,
 width = 200mm,
 pos = htb!,
 center
] {lcccc}{ \tnote[]{\textit{Note}. ML = maximum likelihood, $CI$ = confidence interval, $FPE$ = fungible parameter estimate, $df$ = degrees of freedom, and $\delta$ = perturbation in the scale of $F$. Additionally, $\tilde\epsilon$ = perturbation to the RMSEA and $\delta_F$ = perturbation to $\hat{F}$ by a percentage.}}
  {\FL
	     		& Spirituality $\rightarrow$ Thriving & Spirituality $\rightarrow$ Religiosity	& Religiosity $\rightarrow$  Thriving	& $\delta$\ML
	     ML Estimate	& $0.706$			& $0.617$ 		& $0.252$			&\\
	     $CI$($df=1$)& $[0.526,0.969]$	&				&					& $0.0038$\\
	     $CI$($df=2$)&					&$[0.465,0.773]$&$[-0.123,0.463]$	& $0.0060$\\
	     $FPE_1$ ($\tilde\epsilon_1 = .001$)	& $(0.460,1.144)$	&$(0.448,0.791)$&$(-0.199,0.485)$	& $0.0074$\\
	     $FPE_2$ ($\tilde\epsilon_2 = .00394$)	& $(0.197,2.323)$	&$(0.273,0.960)$&$(-1.481,0.706)$	& $0.0300$\\
	     $FPE_3$ ($\delta_{F_1} = 0.02\hat{F}$)	& $(0.500,1.030)$	&$(0.477,0.760)$&$(-0.076,0.449)$	& $0.0051$\\
	     $FPE_4$ ($\delta_{F_2} = 0.05\hat{F}$)	& $(0.386,1.354)$   &$(0.395,0.852)$&$(-0.458,0.548)$	& $0.0127$\LL}
\end{landscape}
\clearpage

\end{doublespacing}
\clearpage
\clearpage

\clearpage
\usetikzlibrary{shapes}
\begin{figure}[htb!]
\vspace{.55in} \centering
\tikzstyle{e}=[circle, thick, minimum
                size=1cm, draw=black!80,fill=white!20]
\tikzstyle{mv}=[rectangle, thick, minimum
                size=1cm, draw=black!80,fill=white!20]
\tikzstyle{lv}=[ellipse, thick, minimum
                height=1.5cm, draw=black!80,fill=white!20,
                text centered, text width=2cm]
\tikzstyle{v}=[arc, ]
\begin{tikzpicture}
  \path
    (0,10)   node (d1) [e] {$\delta_1$}
    (0,8.5)  node (d2) [e] {$\delta_2$}
    (0,7)    node (d3) [e] {$\delta_3$}
    (5,.0)   node (0)  [] {0}
    (15,2.5) node (02) [] {0}
    (8.5,1)  node (e1) [e] {$\epsilon_1$}
    (11.5,1) node (e2) [e] {$\epsilon_2$}
    (12.5,5)   node (z1) [e] {$\zeta_1$}
    (16.5,7.5) node (z2) [e] {$\zeta_2$}
    (1.5,10)   node (x1) [mv] {$x_1$}
    (1.5,8.5)  node (x2) [mv] {$x_2$}
    (1.5,7)    node (x3) [mv] {$x_3$}
    (5,1.5)    node (x4) [mv] {$x_4$}
    (8.5,3)    node (y1) [mv] {$y_1$}
    (11.5,3)   node (y2) [mv] {$y_2$}
    (15,4)     node (y3) [mv] {$y_3$}
    (5,8.5) node (a) [lv] {Attitude}
    (5,3.5) node (sn)[lv] {Subjective Norm}
    (10,6)  node (i) [lv] {Intention}
    (15,6)  node (t) [lv] {Trying}
    ;
   \path[->, thick]
     (d1) edge  (x1)
     (d2) edge  (x2)
     (d3) edge  (x3)
     (0) edge (x4)
     (a) edge (x1)
     (a) edge (x2)
     (a) edge (x3)
     (a) edge (i)
     (sn) edge (i)
     (i) edge (y1)
     (i) edge (y2)
     (i) edge (t)
     (02) edge (y3)
     (e1) edge (y1)
     (e2) edge (y2)
     (z1) edge (i)
     (z2) edge (t)
   ;
   \draw[<->, thick] (a) to [bend right=30] (sn);
   \draw[->, thick] (sn) -- (x4) node [midway,left]{\small1.00};
   \draw[->, thick] (t) -- (y3) node [midway,left]{\small1.00};
   \draw[->, thick] (a) -- (i)node [midway,above]{\small0.639};
   \draw[->, thick] (a) -- (t) node [midway,above]{\small0.050};
   \draw[->, thick] (i) -- (t) node [midway,above]{\small0.486};
\end{tikzpicture}
\caption{Path Diagram for Example 1}
\label{fig:ex1}
\end{figure}
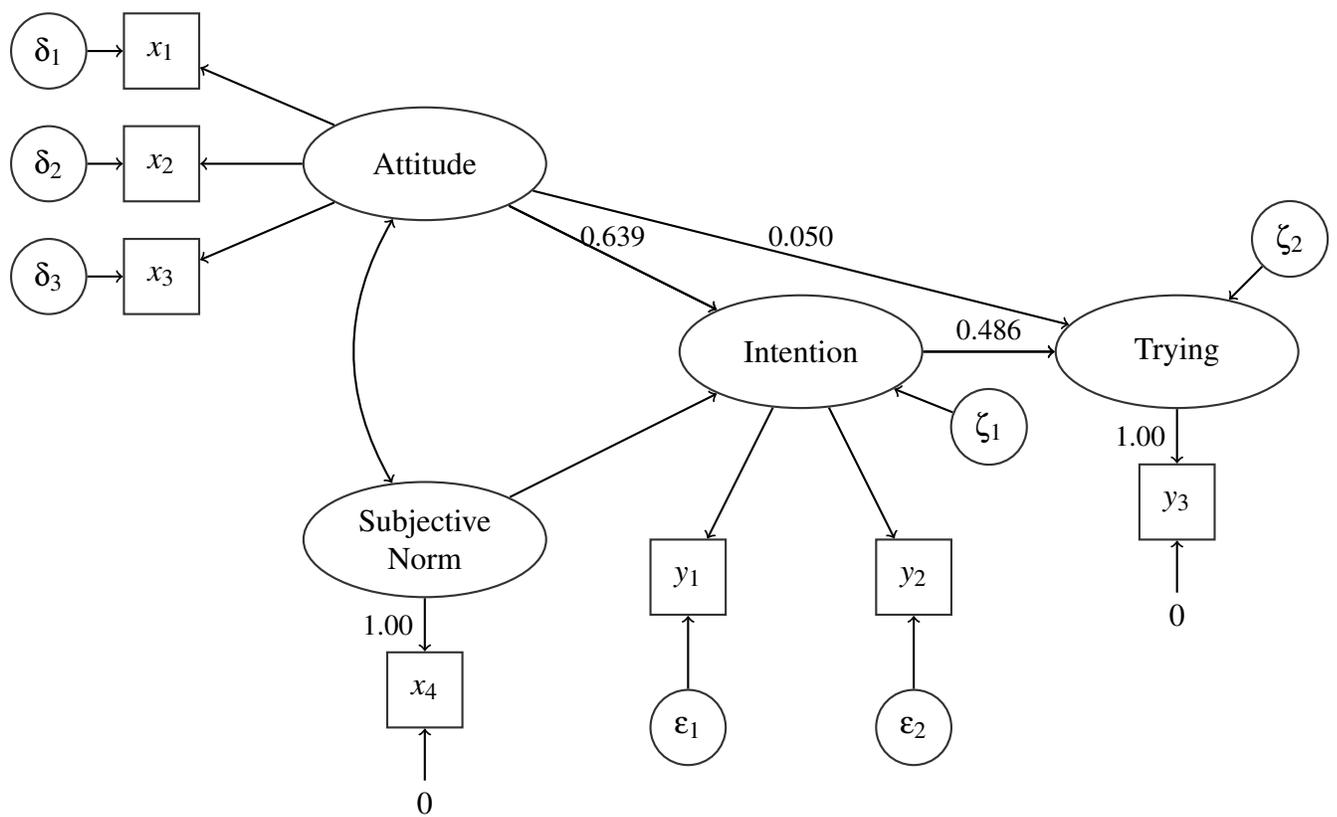
\newpage
\clearpage
\begin{figure}[htb!]
\vspace{.55in} \centering
\tikzstyle{e}=[circle, thick, minimum
                size=1cm, draw=black!80,fill=white!20]
\tikzstyle{mv}=[rectangle, thick, minimum
                size=1cm, draw=black!80,fill=white!20]
\tikzstyle{lv}=[ellipse, thick, minimum
                height=1.7cm, draw=black!80,fill=white!20,
                text centered, text width=2.2cm]
\tikzstyle{v}=[arc, ]
\begin{tikzpicture}
      \path
    (-2.5, 13.5) node (d1) [e] {$\delta_1$}
    (-1, 13.5)   node (d2) [e] {$\delta_2$}
    (0.5, 13.5)  node (d3) [e] {$\delta_3$}
    (2, 13.5)    node (d4) [e] {$\delta_4$}
    (11,11)      node (e1) [e] {$\epsilon_1$}
    (11,9.5)     node (e2) [e] {$\epsilon_2$}
    (11,8)       node (e3) [e] {$\epsilon_3$}
    (11,6.5)     node (e4) [e] {$\epsilon_4$}
    (11,5)       node (e5) [e] {$\epsilon_5$}
    (7,10)     node (z1) [e] {$\zeta_1$}
    (-5.5,8) node (mcgx) [mv] {$x_5$}
    (0,3.5)  node (spx)  [mv] {$x_6$}
    (-2.5,12) node (x1) [mv] {$x_1$}
    (-1, 12) node (x2) [mv] {$x_2$}
    (0.5,12) node (x3) [mv] {$x_3$}
    (2, 12) node (x4) [mv] {$x_4$}
    (9.5,11) node (y1) [mv] {$y_1$}
    (9.5,9.5) node (y2) [mv] {$y_2$}
    (9.5,8) node (y3) [mv] {$y_3$}
    (9.5,6.5) node (y4) [mv] {$y_4$}
    (9.5,5) node (y5) [mv] {$y_5$}
    (0,10) node (pwb) [lv] {Psychological Well-being}
    (6,8) node (ptg)[lv] {Posttraumatic Growth}
    (0,8) node(mcg) [lv]{Meaning in Caregiving}
    (0,6)  node (sp)  [lv] {Spirituality}
;
       \path[->, thick]
      (d1) edge (x1)
      (d2) edge (x2)
      (d3) edge (x3)
      (d4) edge (x4)
      (e1) edge (y1)
      (e2) edge (y2)
      (e3) edge (y3)
      (e4) edge (y4)
      (e5) edge (y5)
     (mcg) edge  (ptg)
     (sp) edge (ptg)
     (pwb) edge (ptg)
     (pwb) edge (x1)
     (pwb) edge (x2)
     (pwb) edge (x3)
     (pwb) edge (x4)
     (ptg) edge (y1)
     (ptg) edge (y2)
     (ptg) edge (y3)
     (ptg) edge (y4)
     (ptg) edge (y5)
     (sp) edge (y5)
     (z1) edge (ptg)
;
   \draw[<->, thick] (pwb) to [bend right=75] (mcg);
    \draw[<->, thick] (pwb) to [bend right=90] (sp);
     \draw[<->, thick] (mcg) to [bend right=75] (sp);
     \draw[->, thick] (mcg) -- (mcgx) node [midway,above]{\small$\sqrt{0.82}$};
     \draw[->, thick] (sp) -- (spx) node [midway,right]{\small$\sqrt{0.93}$};

\end{tikzpicture}
\caption{Path Diagram for Example 2. \textit{Note}. $x_1$ = Self-esteem, $x_2$ = Optimism, $x_3$ = Depression, $x_4$ = Caregiver Burden, $y_1$ = New Possibilities, $y_2$ = Relating to Others, $y_3$ = Personal Strength, $y_4$ = Appreciation of Life, and $y_5$ = Spiritual Change.}
\label{fig:ex2}
\end{figure}
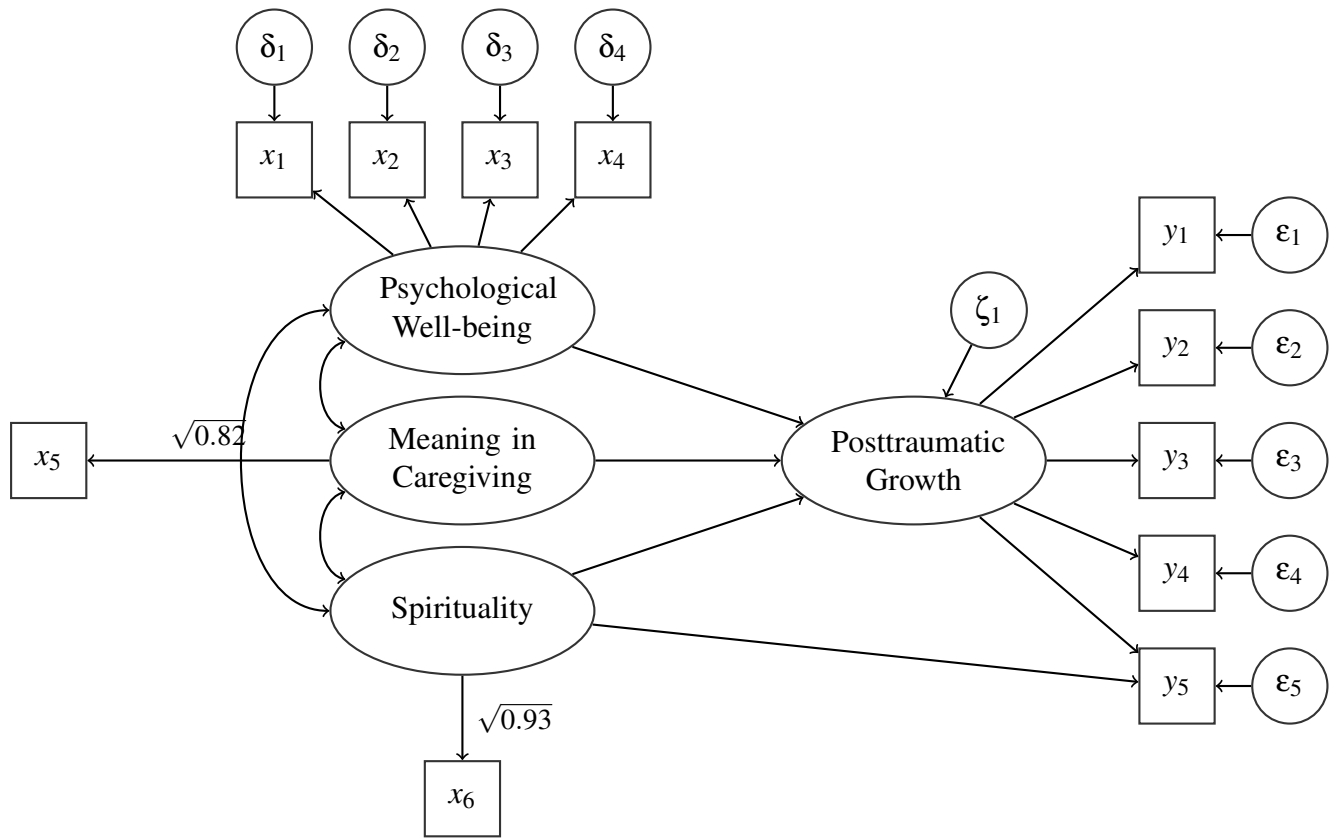
\newpage

\end{document}